# A nearby long gamma-ray burst from a merger of compact objects

E. Troja[1,2,*], C. L. Fryer[3], B. O'Connor[4,5,6,7,*], G. Ryan[8], S. Dichiara[9], A. Kumar[10,11], N. Ito[12], R. Gupta[10,13], R. Wollaeger[3], J. P. Norris[14], N. Kawai[12], N. Butler[2], A. Aryan[10,13], K. Misra[10], R. Hosokawa[12], K. L. Murata[12], M. Niwano[12], S. B. Pandey[10], A. Kutyrev[6,7], H. J. van Eerten[15], E. A. Chase[3], Y.-D. Hu[16], M. D. Caballero-Garcia[16], A. J. Castro-Tirado[16, 17]

[1]Department of Physics, University of Rome "Tor Vergata," Via della Ricerca Scientifica 1, I-00133 Rome, Italy

[2]School of Earth and Space Exploration, Arizona State University, Tempe, AZ 85287, USA

[3]Center for Theoretical Astrophysics, Los Alamos National Laboratory, Los Alamos, NM, 87545, USA

[4]Department of Physics, The George Washington University, 725 21st Street NW, Washington, DC 20052, USA

[5]Astronomy, Physics and Statistics Institute of Sciences (APSIS), The George Washington University, Washington, DC 20052, USA

[6]Department of Astronomy, University of Maryland, College Park, MD 20742-4111, USA

[7]Astrophysics Science Division, NASA Goddard Space Flight Center, 8800 Greenbelt Rd, Greenbelt, MD 20771, USA

[8]Perimeter Institute for Theoretical Physics, 31 Caroline St. N., Waterloo, ON N2L 2Y5, Canada

[9]Department of Astronomy and Astrophysics, The Pennsylvania State University, 525 Davey Lab, University Park, PA 16802, USA

[10]Aryabhatta Research Institute of Observational Sciences (ARIES), Manora Peak, Nainital-263002, India

[11]School of Studies in Physics and Astrophysics, Pandit Ravishankar Shukla University, Chattisgarh 492010, India

[12]Department of Physics, Tokyo Institute of Technology, 2-12-1 Ookayama, Meguro-ku, Tokyo 152-8551, Japan

[13]Department of Physics, Deen Dayal Upadhyaya Gorakhpur University, Gorakhpur 273009, India

[14]Department of Physics, Boise State University, 1910 University Drive, Boise ID 83725-1570 USA

[15]Physics Department, University of Bath, Claverton Down, Bath BA2 7AY, United Kingdom

[16]Instituto de Astrofísica de Andalucía (IAA), CSIC, Granada, Spain.

[17]Unidad Asociada al CSIC Departamento de Ingeniería de Sistemas y Automática, Escuela de Ingeniería Industrial, Universidad de Málaga, Spain

**Gamma-ray bursts (GRBs) are flashes of high-energy radiation arising from energetic cosmic explosions. Bursts of long (>2 s) duration are produced by the core-collapse of massive stars[1], those of short (< 2 s) duration by the merger of compact objects, such as two neutron stars[2] (NSs). A third class of events with hybrid high-energy properties was identified[3], but never conclusively linked to a stellar progenitor. The lack of bright supernovae rules out typical core-collapse explosions[4,5,6], but their distance scales prevent sensitive searches for direct signatures of a progenitor system. Only tentative evidence for a kilonova has been presented[7,8]. Here we report observations of the exceptionally bright GRB211211A that classify it as a hybrid event and constrain its distance scale to only 346 Mpc. Our measurements indicate that its lower-energy (from ultraviolet to near-infrared) counterpart is powered by a luminous (~$10^{42}$ erg s$^{-1}$) kilonova possibly formed in the ejecta of a compact object merger.**

On 11 December 2021 at 13:59:09 Universal Time (UT; hereafter $T_0$), NASA's *Neil Gehrels Swift* observatory (hereafter *Swift*) discovered GRB211211A[9] as an extremely bright burst with a duration of over 50 s (Extended Data Figure 1). The burst was independently observed by the *Fermi*, *INTEGRAL*, and *CALET* satellites. Its optical, ultraviolet (UV) and X-ray counterparts were localized within minutes, close to a nearby galaxy, SDSS J140910.47+275320.8 (G1 in Figure 1), at a distance of 346 Mpc (Methods). Spectroscopic observations of the optical counterpart showed a featureless continuum[10] and did not allow for a direct measurement of the GRB distance scale. However, when combined with the detection of a bright UV counterpart, these observations point to a low-redshift origin for GRB211211A ($z<1.5$ at the 99.9% confidence level; Methods).

Despite the GRB close distance, deep imaging with the *Hubble Space Telescope* (Figure 1) does not detect any underlying host galaxy down to *F160W*> 27.6 AB mag. Several extended objects are visible within 10 arcsec from the GRB position, however their probability of chance superposition is high (>10%; see Methods). The most likely birthsite is in the outskirts of the nearby galaxy G1, at a projected physical offset of 8.00 ± 0.04 kpc from the galaxy's nucleus. This association is also supported a) by probabilistic arguments, being the chance alignment between the GRB and the bright G1 galaxy only 1.4%, b) by the uncommon brightness of the prompt gamma-ray emission, being a total fluence ~3×10$^{-4}$ erg cm$^{-2}$ (15-150 keV), the second highest value ever recorded by *Swift*[11], and c) by the faintness of the X-ray counterpart, as the X-ray flux to gamma-ray fluence ratio, log $f_{X,11hr}/F_\gamma$ ~ -7.9, lies below the typical GRB distribution[12] as expected for an explosion in a rarefied circumburst medium[13,14] (Methods).

The association with a galaxy at 346 Mpc implies that GRB211211A is one of the closest long bursts ever discovered, yet the properties of its gamma-ray emission, such as the negligible temporal lag, short variability timescale, and hard spectrum, do not fit into this class of events (Extended Data Figure 2). These are distinctive features of short bursts and classify GRB211211A as a hybrid event, analogous to GRB060614[3]. In addition to its prompt gamma-ray phase, several lines of evidence differentiate GRB211211A from canonical long GRBs. The GRB does not lie in a star-forming region (Methods) and late-time optical imaging rules out any bright supernova (SN) at its location (Extended Data Figure 4): as the dust content along the line of sight is negligible, a luminous SN similar to SN 1998bw[15] is excluded out to $z$~0.8. A faint and short-lived SN similar to SN2008ha[6] is also ruled out by the optical limits. The GRB location and the global properties of its host galaxy provide indirect evidence for a stellar progenitor different from a collapsing massive star and are instead consistent with a compact binary merger (Methods).

The unambiguous proof of a compact object binary merger comes either from its gravitational wave signal[2] or from its kilonova, a short-lived glow of quasi-thermal radiation powered by the radioactive decay energy of heavy nuclei[16], produced in the merger ejecta via rapid neutron capture process (r-process). The first known kilonova was AT2017gfo, characterized by an early (<12 hrs) UV/optical peak[17] followed by a longer-lasting infrared signal[18,19,20,21]. We find that a similar component is identified in the UV/optical/infrared (UVOIR) counterpart of GRB211211A, providing us with the direct link to compact binary mergers.

The multi-wavelength emission that follows a GRB is the superposition of multiple components. The dominant one is usually the afterglow, a broadband synchrotron radiation emitted by a population of electrons shock-accelerated by the GRB outflow[22]. We use the X-ray data to probe the contribution of this non-thermal component. The X-ray spectrum is well-described by a power-law with slope $\beta_X \sim 0.5$ and negligible absorption along the line of sight. When extrapolated to lower energies, this model roughly matches the observed optical fluxes at $T_0+1$ hr and shows no evidence for an additional component at this time. However, at later times, the multi-frequency spectral energy distribution (Figure 2) identifies emission in excess of the standard afterglow: the UVOIR counterpart is consistently brighter than the extrapolation of the non-thermal power-law, and is characterized by a steeper spectral index $\beta_{UVOIR} > 2$ for $t > 1$ d. Its spectral peak lies in the UV range (*u* band, observer-frame) at $T_0+0.2$ d and then progressively cools down to near-infrared (nIR) wavelengths (*K* band at $\sim T_0+4$ d).

We rule out that a reverse-shock (RS) powered afterglow or a supernova (SN) onset could explain this low-energy component. The former arises within the GRB outflow and is characterized by an optical rebrightening peaking from a few seconds to ~1 hr after the burst[23,24]. However, RS emission quickly cools off and shifts to the radio band, typically within the first day after the burst.

This is not consistent with the observed SED evolution. Moreover, a low nickel-yield explosion would also produce a short-lived UV/optical flare powered by shock heating in the SN blast-wave[25]. We studied a broad range of collapsar-associated SNe varying nickel-yields, stellar properties and explosion energies. Although this model explains the lack of a bright SN at late times and can reproduce the basic features of the early optical emission such as the bolometric luminosity and photospheric radii (Extended Data Table 1), the predicted spectrum is too hard (Extended Data Figure 6): UV emission dominates and we cannot reproduce the bright and long-lived nIR emission without the addition of a second, neutron rich outflow (see Methods).

After subtracting the afterglow contribution from the data, we find that the UVOIR excess is well described by a thermal spectrum and that the best-fit parameters point to a hot ($T$~15,000 K, rest-frame) fireball in rapid expansion with apparent velocity $v \gtrsim 0.5\ c$. These properties do not match neither those of optical transients from white dwarf mergers (Supplementary Methods) nor those of a thermal dust echo[26]. Instead, the luminosity, temperature and emitting radius of this thermal component display a striking resemblance to AT2017gfo[27] (Figure 2), and we interpret it as the kilonova emission associated with GRB211211A. A kilonova in GRB211211A and consequently its association with a compact binary merger tie the lack of SN, the GRB environment, and the evolution of its UVOIR counterpart in a coherent explanation.

Our dataset allows us to probe the earliest phases of the kilonova onset, not observed in the case of AT2017gfo. Although the broadband emission is initially dominated by the non-thermal afterglow, evidence for a thermal component is found as early as $T_0 + 5$ hr. Figure 3 shows the different behaviors of the X-ray and UVOIR counterparts. The latter requires an additional component, which we model using simulated kilonova light curves[28] with wind ejecta mass $M_w$ in the range 0.01 - 0.1 $M_\odot$, and dynamical ejecta mass $M_d$~0.01-0.03 $M_\odot$. The ejecta velocity and kilonova bolometric luminosity, $L_{bol}$ ~ 3×10$^{42}$ erg s$^{-1}$ (isotropic equivalent) inferred at early times, are challenging to reproduce with purely radioactive-powered models[28], even when accounting for different density profiles and the larger projected area along the polar axis[29] (Methods). We therefore explore alternative models in which the merger ejecta is re-energized by a central engine or modified by the interaction with the GRB jet. The former group of models, envisioning either a highly magnetized NS or fallback accretion onto the central black hole (BH), is often invoked to explain a long-lasting gamma-ray emission[30,31]. However, an active engine would leave observable imprints on the kilonova light[32], which are not consistent with its timescales (too early) or colors (too red) (Extended Data Figure 6).

We therefore consider a model in which jet-ejecta interactions shape the observed emission. A relativistic jet is present in both GRB211211A and GRB170817A and its effects may explain their similar kilonova evolution. As the jet propagates through the massive (≳ 0.01 $M_\odot$) cloud of radioactive ejecta, it heats and partially disrupts its density structure, carving a funnel of low-opacity low-density material along the polar axis[32]. By exposing the inner, hotter surface of the ejecta, an energetic ($E_{\gamma,iso}$~6×10$^{51}$ erg) GRB jet makes the kilonova emission both bluer and brighter[33] for an observer close to its axis. Shock-heating may also contribute to distribute the energy. Viceversa, the ejecta imparts a wide angular structure on the GRB jet before it breaks out[34,35]. High latitude emission from the jet wings arrives later because of the longer path that photons travel and may produce a low-luminosity fast-fading X-ray transient[36] consistent with the observed X-ray behavior. This feature may become visible in the case of a "naked" structured GRB jet expanding into a low-density circumburst medium, such as GRB211211A.

We conclude that, although the long duration of the prompt phase challenges our understanding of compact binary merger models, a merger progenitor naturally explains all the other observed features of GRB211211A. At 346 Mpc, this GRB lies within the distance horizon of forthcoming gravitational wave (GW) observing runs[37] and, had the GW network been online at the time of the burst, this event would have likely resulted in a joint detection of GWs and electromagnetic (EM) radiation. We note that its EM properties are very different from the multi-messenger transient GW170817: whereas the EM counterparts of GW170817[2,17,20] would be challenging to localize beyond ~150 Mpc, GRB211211A would be visible out to $z$~1 by most space-borne gamma-ray detectors. Moreover, rapid X-ray and UV/optical follow-up of GW sources would detect its counterpart out to $z$~0.2 assuming a sensitivity comparable to *Swift*.

To determine the rate of hybrid GRBs, we examine the *Swift* GRB catalog[11] in search of bursts like GRB211211A and GRB060614. At large distances ($z$~1), their classification would rely solely on the high-energy properties which point to regular bursts of long duration (Extended Data Figure 7). Without a systematic study of GRB lags, spectra and durations, it is not possible to assess the total number of hybrid bursts thus far detected. Therefore, we turn to lower redshifts where a clear observational signature of these events is the lack of a SN. SNe associated with GRBs[1] peak between $M_V$~-18.5 mag and $M_V$~-20 mag, and sensitive SN searches are regularly undertaken for GRBs within a redshift $z$<0.3, which we identify as the maximum distance for a homogeneous identification. We review the entire sample of *Swift* bursts with duration >2 s and a putative host galaxy at $z$<0.3 and find a total of 20 GRBs in 17 years of mission (2005-2021). Of these, 8 are associated with a SN, 3 have no meaningful constraints, and 9 have deep limits on any accompanying SN. The chance alignment between a bright galaxy and an afterglow with sub-arcsecond localization is typically <1% (ref. 11), thus it is unlikely that all nine bursts are distant background objects. Furthermore, four of them (GRB060614, GRB060505, GRB191019A and GRB211211A) have UV counterparts constraining their distance scale[5,38,39]. We conclude that some of these long duration bursts are physically associated with a low-redshift galaxy and lack a SN, forming a new class of hybrid GRBs produced by compact binary mergers. After accounting for instrumental effects (Supplementary Methods), we derive a volumetric all-sky rate of 0.04-0.8 Gpc$^{-3}$ yr$^{-1}$ (68% c. l.), lower than the observed rate of short GRBs[40]. The true rate of events depends on the unknown beaming factor $f_b$ of these outflows. Assuming similar jet properties to short GRBs[41], hybrid long duration bursts may represent ~10% (0.8%-26% at the 68% c.l.) $f_{b,short}/f_{b,hybrid}$ of the population of EM counterparts to GW sources caused by compact binary mergers.

## References

1. Woosley, S. E. & Bloom, J. S. The supernova gamma-ray burst connection. *Annu. Rev. Astron. Astrophys.* **44**, 507–556 (2006)
2. Abbott, B. P. et al. Gravitational Waves and Gamma-Rays from a Binary Neutron Star Merger: GW170817 and GRB 170817A. *Astrophys. J.* **848**, L13 (2017)
3. Gehrels, N. et al. A new γ-ray burst classification scheme from GRB060614. *Nature* **444**, 1044-1046 (2006)
4. Della Valle et al. An enigmatic long-lasting γ-ray burst not accompanied by a bright supernova. *Nature* **444**, 1050-1052 (2006)
5. Gal-Yam, A. et al. A novel explosive process is required for the γ-ray burst GRB 060614. *Nature* **444**, 1053-1055 (2006)
6. Valenti, S. et al. A low-energy core-collapse supernova without a hydrogen envelope. *Nature*. **459**, 674-677 (2009)
7. Yang, B. et al. A possible macronova in the late afterglow of the long-short burst GRB 060614. *Nat. Commun.* **6**, 7323 (2015)
8. Jin, Z.-P. et al. The Light Curve of the Macronova Associated with the Long-Short Burst GRB 060614. *Astrophys. J.* **811**, L22 (2015)
9. D'Ai, A., et al., GRB 211211A: Swift detection of a bright burst. *GCN Circ.* **31202** (2021)
10. Malesani, D. B., et al. GRB 211211A: NOT optical spectroscopy. *GCN Circ.* **31221** (2021)
11. Lien, A., et al. The Third Swift Burst Alert Telescope Gamma-Ray Burst Catalog. *Astrophys. J.* **829**, 7 (2016)
12. O'Connor et al. A deep survey of short GRB host galaxies over z~0-2: implications for offsets, redshifts, and environments. *Mon. Not. R. Astron. Soc.* doi:10.1093/mnras/stac1982 (2022)
13. Kumar, P. & Panaitescu, A., Afterglow Emission from Naked Gamma-Ray Bursts. *Astrophys. J.* **541**, L51-L54 (2000)

14. O'Connor, B. et al. Constraints on the circumburst environments of short gamma-ray bursts. *Mon. Not. R. Astron. Soc.* **495**, 4782-4799 (2020)
15. Galama, T. J. et al. An unusual supernova in the error box of the γ-ray burst of 25 April 1998. *Nature* **395**, 670-672 (1998)
16. Li, L.-X. & Paczyński, B. Transient Events from Neutron Star Mergers. *Astrophys. J.* **507**, L59-L62 (1998)
17. Evans, P. A. et al. Swift and NuSTAR observations of GW170817: Detection of a blue kilonova. *Science* **358**, 1565-1570 (2017)
18. Pian, E. et al. Spectroscopic identification of r-process nucleosynthesis in a double neutron-star merger. *Nature* **551**, 67-70 (2017)
19. Smartt, S. J. et al. A kilonova as the electromagnetic counterpart to a gravitational-wave source. *Nature* **551**, 75-79 (2017)
20. Troja, E. et al. The X-ray counterpart to the gravitational-wave event GW170817. *Nature* **551**, 71-74 (2017)
21. Kasliwal, M. M. et al. Spitzer mid-infrared detections of neutron star merger GW170817 suggests synthesis of the heaviest elements. *Mon. Not. R. Astron. Soc.* **510**, L7-L12 (2022)
22. Kumar, P. & Zhang, B. The physics of gamma-ray bursts & relativistic jets. *Phys. Rep.* **561**, 1–109 (2015)
23. Dichiara, S. et al. The early afterglow of GRB 190829A. *Mon. Not. R. Astron. Soc.* **512**, 2337-2349 (2022)
24. Japelj, J. et al. Phenomenology of Reverse-shock Emission in the Optical Afterglows of Gamma-Ray Bursts. *Astrophys. J.* **785**, 84 (2014)
25. Fryer, C. L., Hungerford, A. L. & Young, P. A. Light-Curve Calculations of Supernovae from Fallback Gamma-Ray Bursts. *Astrophys. J.* **662**, L55-L58 (2007)
26. Waxman, E., Ofek, E. O. & Kushnir, D. Strong NIR emission following the long duration GRB 211211A: dust heating as an alternative to a kilonova. Preprint at https://arxiv.org/abs/2206.10710 (2022)
27. Waxman E., Ofek E. O., Kushnir D., Gal-Yam A., Constraints on the ejecta of the GW170817 neutron star merger from its electromagnetic emission. *Mon. Not. R. Astron. Soc.* **481**, 3423 (2018)
28. Wollaeger, R. T. et al., A Broad Grid of 2D Kilonova Emission Models. *Astrophys. J.* **918**, 10 (2021)
29. Gompertz, B. P., O'Brien, P. T., Wynn, G. A. & Rowlinson, A. et al. Can magnetar spin-down power extended emission in some short GRBs?. *Mon. Not. R. Astron. Soc.* **431**, 1745-1751 (2013)
30. Rosswog S., Fallback accretion in the aftermath of a compact binary merger. *Mon. Not. R. Astron. Soc.* **376**, L48 (2007)
31. Wollaeger, R. T. et al., Impact of Pulsar and Fallback Sources on Multifrequency Kilonova Models. *Astrophys. J.* **880**, 22 (2019)
32. Korobkin O., et al., Axisymmetric Radiative Transfer Models of Kilonovae, *Astrophys. J.* **910**, 116 (2021)
33. Nativi L., et al. Can jets make the radioactively powered emission from neutron star mergers bluer? *Mon. Not. R. Astron. Soc.* **500**, 1772-1783 (2021)
34. Bromberg, O., Nakar, E., Piran, T. & Sari, R. The Propagation of Relativistic Jets in External Media. *Astrophys. J.* **740**, 100 (2011)

35. Lazzati, D. et al. Late Time Afterglow Observations Reveal a Collimated Relativistic Jet in the Ejecta of the Binary Neutron Star Merger GW170817. *Phys. Rev. Lett.* **120**, 241103 (2018)
36. Ascenzi, S. et al., High-latitude emission from the structured jet of γ-ray bursts observed off-axis. *Astron. Astrophys.* **641**, A61 (2020)
37. Petrov, P. et al. Data-driven Expectations for Electromagnetic Counterpart Searches Based on LIGO/Virgo Public Alerts. The *Astrophys. J.* **924**, 54 (2022)
38. Ofek, E. O. et al. GRB 060505: A Possible Short-Duration Gamma-Ray Burst in a Star-forming Region at a Redshift of 0.09. *Astrophys. J.* **662**, 1129-1135 (2007)
39. LaPorte, S. J., Simpson, K. K. & Swift/UVOT Team GRB 191019A: Swift/UVOT Detection. *GCN Circ.* **26053**, 1 (2019)
40. Wanderman, D. & Piran T. The rate, luminosity function and time delay of non-Collapsar short GRBs. *Mon. Not. R. Astron. Soc.* **448**, 3026-3037 (2015)
41. Troja, E. et al. An Achromatic Break in the Afterglow of the Short GRB 140903A: Evidence for a Narrow Jet. *Astrophys. J.* **827**, 102 (2016)

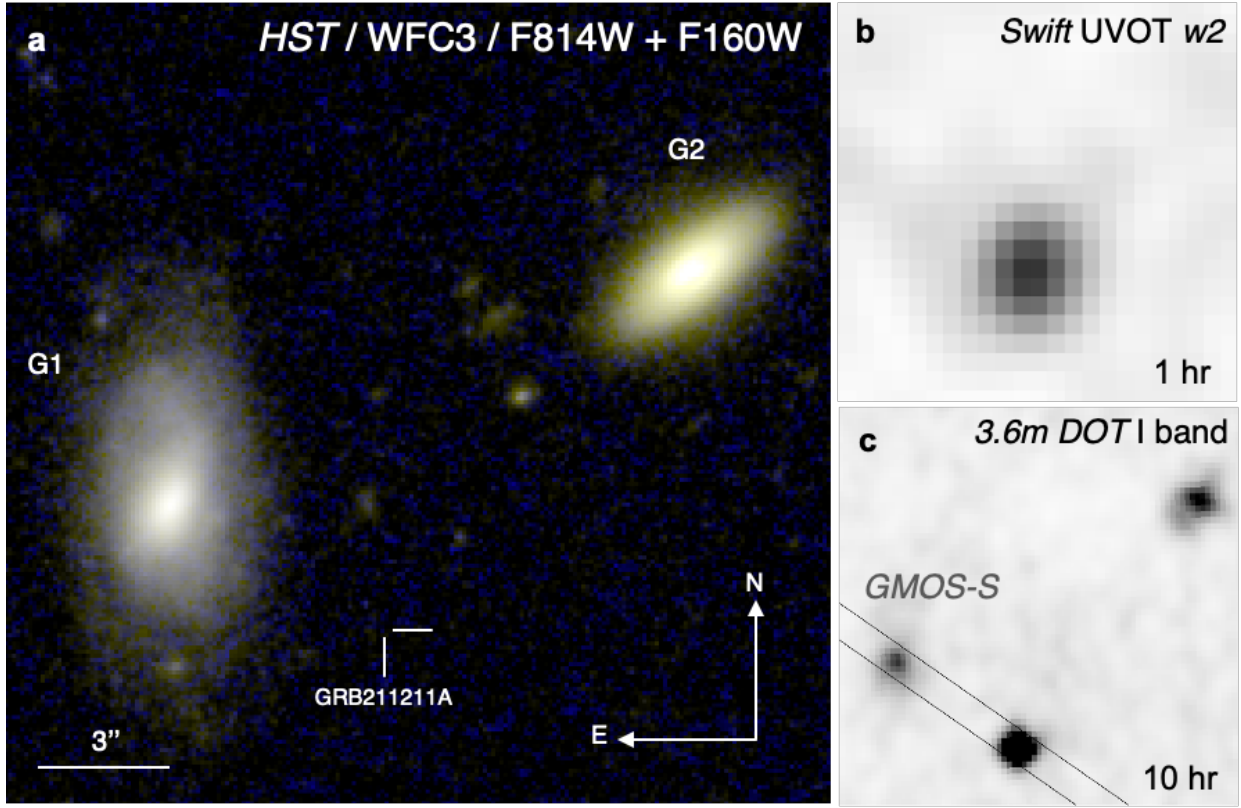

**Figure 1 - The field of GRB211211A**

This false color image (a) combines optical (blue) and near-infrared (red+green) *HST* observations of GRB211211A, carried out in April 2022 (~4 months after the burst). Two bright galaxies (G1 at $z\sim0.0762$, and G2 at $z\sim0.4587$) and several fainter ones are visible, but no source is detected at the location of GRB211211A. The most likely host galaxy is G1, a low mass late-type galaxy. The projected physical offset between the burst and the galaxy's center is ~8 kpc, one of the largest ever measured for a long burst. The same field is shown in the UV *w2* filter (b) and optical I filter (c) at 1 hr and 10 hr after the burst, respectively. The solid lines show the slit position used for optical spectroscopy with *Gemini*/GMOS-S. The bright UV counterpart rules out a high-redshift origin, whereas its rapid reddening is consistent with the onset of a kilonova.

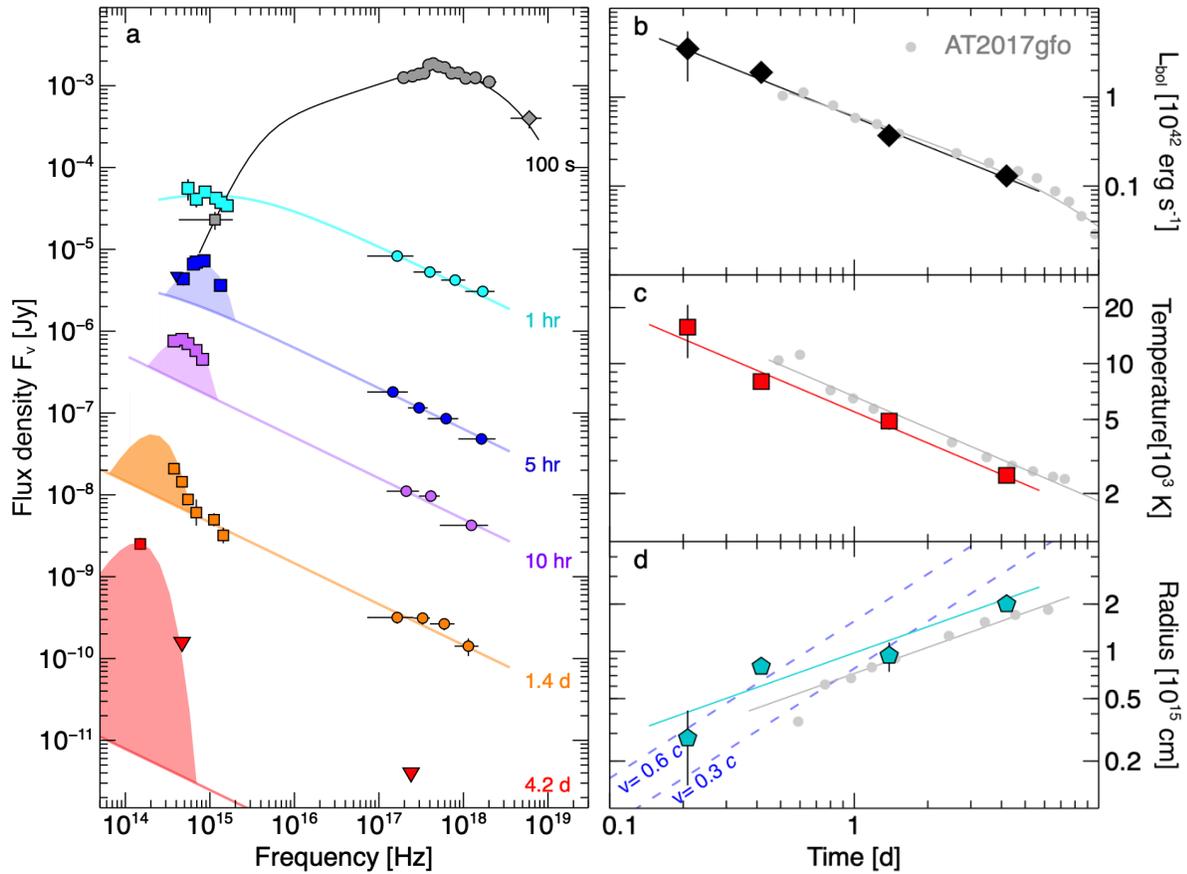

**Figure 2 - Spectral evolution of the GRB afterglow and kilonova**

The spectral energy distribution (a) combines gamma-ray (diamonds), X-rays (circles) and UVOIR (squares) data at different times, as indicated by the labels. It shows that non-thermal radiation (solid line) dominates at early times and at higher energies. At lower energies, we identify the emergence of a thermal component peaking at blue wavelengths at 5 hr, and rapidly shifting toward redder colors. Error bars represent $1\sigma$; upper limits (downward triangles) are $3\sigma$. For plotting purposes, each epoch was rescaled by the following factors (from top to bottom): 1, 1, $10^{-0.8}$, $10^{-1.6}$, $10^{-2.4}$, $10^{-3.2}$. The bolometric luminosity (b), temperature (c), and emitting radius (d) of the thermal component are similar to AT2017gfo[27] (gray circles). Solid lines show the best fit power-law models to the dataset. Dashed lines in panel d show the predicted radius for constant expansion velocities of $0.3\,c$ and $0.6\,c$.

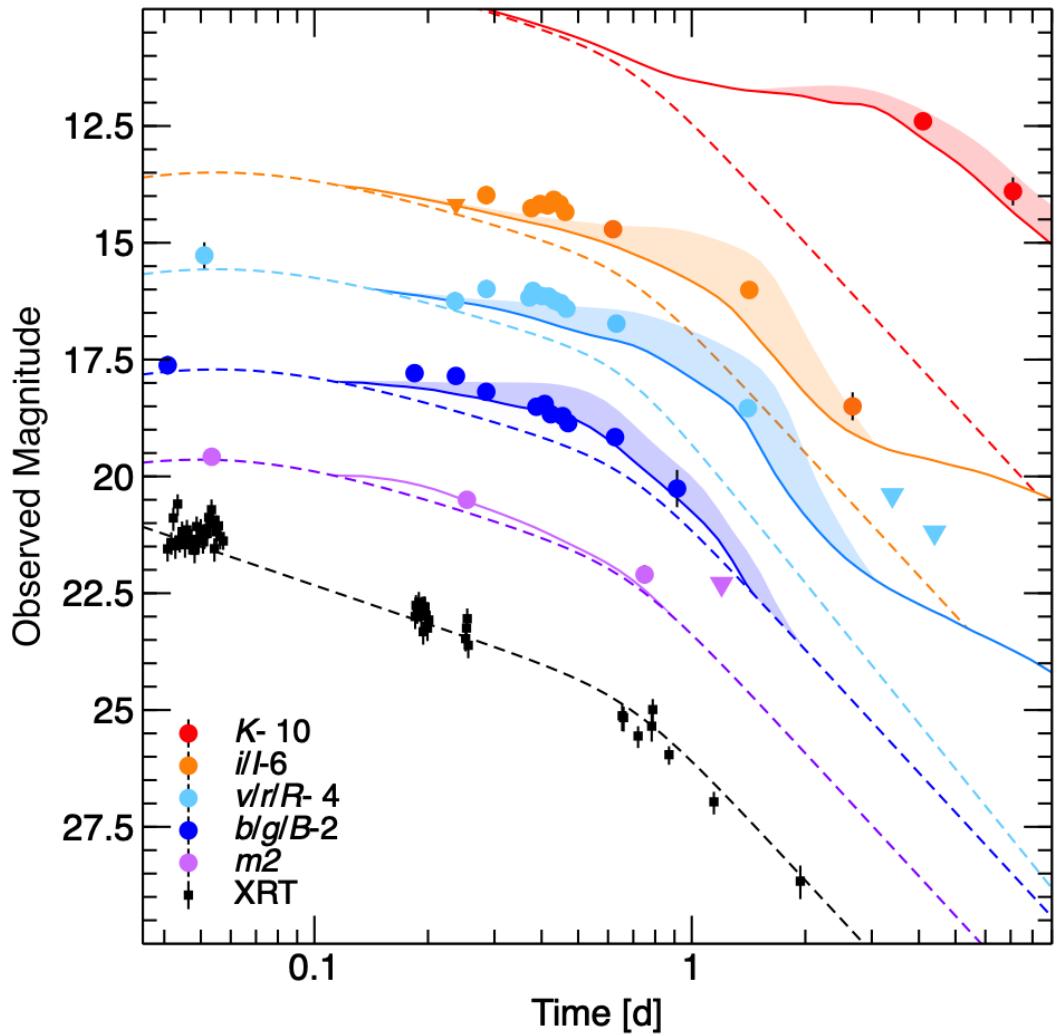

**Figure 3 - A kilonova in the long GRB211211A**

Multi-color light curves in X-rays, UV (*uvm2*), optical (*BRI*) and infrared (*K*) are compared to models' predictions of a kilonova (solid line) in addition to the non-thermal emission (dashed line). The shaded area shows the range of possible fluxes reproduced by kilonova simulations with wind mass $M_w$ between 0.01 $M_\odot$ (lower bound) and 0.1 $M_\odot$ (upper bound), and dynamical ejecta mass $M_d$ between 0.01 $M_\odot$ (lower bound) and 0.03 $M_\odot$ (upper bound). Error bars represent $1\sigma$; upper limits (downward triangles) are $3\sigma$.

## Methods

**Classification of GRB211211A**

GRBs are classified based on the properties of their prompt gamma-ray phase. The prompt emission of GRB211211A (Extended Data Figure 1) displays three different episodes: a weak precursor, a bright multi-peaked main burst, and a highly variable temporally extended emission. The time intervals for spectral and temporal analysis were selected to characterize them separately. *Swift* and *Fermi* data were processed using HEASOFT v.6.30. Spectra were extracted from the *Fermi* Gamma-ray Burst Monitor data and fit within XSPEC[42]. The temporal properties were derived from the *Swift* BAT light curves using well-established techniques[43,44].

The precursor phase has a short duration of 0.15 s, a soft spectrum peaking at ~75 keV, a minimum variability timescale of 21±4 ms, and a significant lag $\tau_{31}= 16^{+4}_{-3}$ ms between the temporal structures observed in the 50-100 keV (band 3) and in the 15-25 keV (band 1) energy bands, respectively. At 346 Mpc, the measured flux of 8x10$^{-7}$ erg cm$^{-2}$ s$^{-1}$ (10 - 1,000 keV) corresponds to a luminosity of only ~10$^{49}$ erg s$^{-1}$.

After a 1 s period of quiescence, we detect the onset of the main prompt emission, which consists of multiple overlapping peaks lasting for approximately 10 s. The time-averaged spectrum peaks at 750±10 keV, the minimum variability timescale is 14±5 ms, and the temporal lag is negligible with $\tau_{31}= -0.9^{+2.8}_{-2.6}$ ms. The total fluence measured during this episode is ~3.7×10$^{-4}$ erg cm$^{-2}$ (10 - 1,000 keV), one of the highest ever measured for a GRB. However, at 346 Mpc the total isotropic-equivalent gamma-ray energy $E_{\gamma,iso}$ would be ~5×10$^{51}$ erg within the typical GRB range[45].

A brief (3 s) period of low-level persistent emission precedes the onset of a long-lasting tail. The time-averaged spectrum of the extended emission has a soft peak of 52±2 keV, the minimum variability timescale is 42 ±9 ms, and the lag, $\tau_{31}= 7^{+3}_{-2}$ ms, is positive. The total fluence is ~5×10$^{-5}$ erg cm$^{-2}$ (10 - 1,000 keV), corresponding to $E_{\gamma,iso}$ ~7×10$^{50}$ erg.

We compare the properties of the main prompt emission to the population of GRBs using four classifiers: the duration/hardness-ratio diagram[46], the lag-luminosity relation[43], the variability timescale[44], and the Amati correlation[45] (Extended Data Figure 2). Similar to GRB060614[3], GRB211211A shows characteristics that are intermediate between the two main GRB classes: the traditional classification based on duration and hardness ratio places this event in the class of long GRBs, however its other properties fit within the class of short bursts. Its hybrid nature does not allow us to unambiguously link it to a progenitor system based solely on its high-energy properties.

**The GRB environment and its host galaxy**

The GRB environment typically offers stringent, albeit indirect, evidence of its progenitor system. In the case of GRB211211A, no underlying host galaxy is detected in late-time *HST* imaging (Figure 1). By planting artificial sources with an exponential disk profile and different brightness, we derive an upper limit of *F814W*>26.5 AB mag and *F160W*>27.6 AB mag. Since no coincident galaxy is found, we analyze the GRB field to search for its most likely host. We identify 7 galaxies within 10'' from the GRB position (see Figure 1): G1 with *r* = 19.50 +/- 0.02 mag at an offset of 5.55'' +/-0.03'', G2 with *r* = 20.88 +/-0.05 mag at an offset of ~10'', and 5 faint (*r* > 26 AB mag) extended objects at an offset between 2.5'' and 10''. By using the galaxy's number counts in r-band[47], we derive a chance alignment of 1.4% for G1, >10% for G2, and >40% for the other faint galaxies. Therefore, probabilistic arguments favor the association between GRB211211A and G1. We note that the probability threshold adopted to associate a galaxy to a GRB is generally >1%, meaning that G1 with $P_{cc}$~1.4% would be considered as the most likely host by any previous studies of GRB galaxies[12,47]. Moreover, in our spectroscopic observations we find no evidence for any emission lines at the GRB position down to $> 2 \times 10^{-17}$ erg cm$^{-2}$ s$^{-1}$ Å$^{-1}$ in the range 4,800-6,100 Å. Using [OII] 3727 and H$\beta$ as indicators of unobscured star formation[48], we place an upper limit on the star formation rate SFR<1 $M_\odot$ yr$^{-1}$ for $z$<0.65. This corresponds to the median SFR of long GRB hosts[49] at $z < 1$, providing additional constraints on any possible underlying galaxy.

The spectrum of G1 shows several emission lines including H$\alpha$, [NII], and [SII] at a common redshift of $z$=0.0762 +/- 0.0003, consistent with the previous report[9] based on data from the Nordic Optical Telescope (NOT). Assuming a $\Lambda$CDM cosmology[50] with H$_0$=69.8 km Mpc$^{-1}$ s$^{-1}$, we find a luminosity distance d$_L$ = 346 Mpc, and a distance modulus $\mu$ = -37.7 mag. Using the host galaxy photometry (Supplementary Table 1), we compute a rest-frame absolute *B*-band magnitude of M$_B\approx$-17.6 AB mag, corresponding to L$_B\approx$0.1L$_*$ when compared to the galaxy luminosity function[51] at a similar redshift (0.05<$z$<0.2).

The brightness ($L_{H\alpha}$~10$^{40}$ erg s$^{-1}$) and relative ratio of these lines (log [NII]/H$\alpha$ ~ -0.7) point to a star-forming galaxy with SFR~0.05 $M_\odot$ yr$^{-1}$ and sub-solar metallicity 12 + log O/H ~ 8.4. We also find evidence for weak [Mg I $\lambda$5175Å] absorption at ~5567 Å, indicative of an evolved stellar population, although this feature is affected by a nearby sky line.

We model the galaxy's surface brightness using GALFIT[52]. A good description ($\chi_\nu^2$~1.03) of its morphology is obtained by including two Sersic profiles with index *n*=1, one with half-light radius R$_{e,1}$~2.15 arcsec (*F814W*; ~3.1 kpc at $z$=0.076) and another one with R$_{e,2}$~0.5 arcsec (*F814W*; ~0.7 kpc at $z$=0.076) to model the central bar. Similar results are obtained on the *F160W* image with R$_{e,1}$~2.34 arcsec and R$_{e,2}$~0.64 arcsec. The half-light radius $r_{50}$ ~ 1.1 arcsec obtained through Source Extractor is given by the weighted average of these two components.

The galaxy's global properties were determined by modeling its SED (Supplementary Table 1) with Prospector[53], adopting the same settings used for GRB host galaxies[12,54]. We derived a stellar mass of $M = 0.9^{+0.2}_{-0.4} \times 10^9 \, M_\odot$, a star-formation rate SFR=$(0.06 \pm 0.02) \, M_\odot$ yr$^{-1}$, a low dust content $A_V = 0.09^{+0.08}_{-0.06}$ mag, and a mass-weighted stellar age $\tau = 5^{+2}_{-3}$ Gyr (Extended Data Figure 5). When compared to the sample of long GRBs, the properties of the host of GRB211211A are not unprecedented but extremely uncommon. The inferred SFR lies in the bottom 10% of the observed distribution, leading to an unusually low specific SFR, sSFR≈0.06 Gyr$^{-1}$. This value is below the main sequence of star forming galaxies[55], indicating that G1 may be migrating to a quiescent phase. This differs from the typical environment of long GRBs at both high and low-redshifts: for comparison, nearby events such as GRB060218 and GRB100316D were associated with sSFR≈4 Gyr$^{-1}$ and sSFR≈0.2 Gyr$^{-1}$, respectively[56,57]. Dissimilarities with the class of short GRBs also exist: the stellar mass lies at the bottom 10% of both short GRB and SN Ia host galaxies[58,59], as for GRB060614 which was hosted by a dwarf galaxy[5].

**Spectral Energy Distribution**

The spectral energy distribution (SED) of the GRB counterpart at different times is shown in Figure 2. These epochs were selected to maximize simultaneous multi-wavelength coverage. When needed, the data were rescaled to a common epoch using the best fit temporal model.

In the first epoch at $T_0+100$ s, the X-ray emission is characterized by a flat spectral index $\beta_X= 0.00$ +/- 0.03. A spectral break is required above ~10 keV to account for the lower flux and soft spectral index, $\beta_{BAT}$~2, measured in the hard X-ray band. In addition, the high X-ray-to-optical flux ratio, $F_X/F_O$≈100, requires a turn-over to a steep spectrum between the X-ray and optical band. These properties are consistent with self-absorbed synchrotron radiation in the fast cooling regime. The location of a self-absorption frequency, $\nu_a$~10 eV, indicates a compact emitting region[60] with radius $R$≈$10^{13}$ $(\Gamma/300)^{3/4}$ cm, where $\Gamma$ is the outflow bulk Lorentz factor. This radius is typical of dissipation processes within the GRB outflow, indicating that at ~$T_0$+100 s the prompt phase is still dominant at both X-ray and optical wavelengths.

In the second epoch at $T_0$+1 hr, the GRB counterpart displays blue colors with a spectral index $\beta_O$ = 0.23±0.10 in the UV and optical bands. At X-ray energies the spectrum, extracted between 3 ks and 5 ks, has a slope of $\beta_X$ = 0.50±0.05. This index points to synchrotron radiation in the slow cooling regime, in which the cooling frequency is $\nu_c$ > 10 keV and the synchrotron frequency is $\nu_m \lesssim$ 1 eV. In this case, the X-ray spectral slope is related to the energy distribution of the emitting electrons, $N(E) \propto E^{-p}$ with $p$ = 2 $\beta_X$ + 1 = 2.0±0.1. This is a fundamental constraint to the long-term afterglow evolution. The steepest spectral slope explained by this model is $p/2 \approx$ 1.05, and only for energies above $\nu_c$. Therefore, the UVOIR and X-ray non-thermal afterglows are bound to remain on the same spectral segment over the time span of our observations.

Starting from ~$T_0$+5 hr, a simple non-thermal spectrum can no longer reproduce the broadband emission. An UVOIR excess is detected at all epochs. It is characterized by a narrow spectral shape peaking in the *u* band, well described by a blackbody function with temperature $T$~16,000 K (rest-frame) and a luminosity $L_{bol} \approx$ (3.5±2.0) × $10^{42}$ erg s$^{-1}$. We therefore fit each SED epoch with a blackbody (UVOIR) plus power-law (X-ray) model, and derive the total integrated blackbody luminosity, its temperature and radius as a function of time (Figure 2 and Extended Data Table 1). The luminosity is better constrained in our second epoch at $T_0$ + 10 hr, $L_{bol}$ = (1.90 ± 0.15) × $10^{42}$ erg s$^{-1}$ and is seen to decrease in time following a power-law $\propto t^{-0.95}$, consistent with the evolution of AT2017gfo[27].

**GRB distance scale**

We investigate the joint X-ray/UV/optical SED at 1 hr to place a direct upper limit on the GRB distance scale. UVOT spectra were created with the tool *uvot2pha* using the same source and background regions selected for photometry. We adopt a power-law model and include the effects of absorption, dust reddening, and intergalactic medium attenuation as implemented in the XSPEC models *zphabs*, *zdust*, and *zigm*. The Galactic absorption was fixed to $N_H = 1.76 \times 10^{20}$ cm$^{-2}$ and the reddening at *E(B-V)* = 0.015 mag. All other parameters were left free to vary. We increase the redshift from 0 to 2.5 in steps of 0.1 and find the best fit model by minimizing the Cash statistics, recording its value at each step. Based on the variations of the test statistics, we derive an upper limit of *z* < 2.3 (99.9% confidence level) from the UV/optical data, and *z*< 1.5 (99.9% confidence level) from the joint X-ray/UV/optical fit. Within this range of distances, a typical GRB host galaxy would be visible in deep *HST* observations (Extended Dta Figure 3). By imposing the redshift of the putative host galaxy G1, *z*~0.0762, we find no evidence for any dust extinction or absorption at the GRB site with 3 σ upper limits of *E(B-V)$_z$* < 0.005 mag and $N_{H,z}$ < 9 × 10$^{19}$ cm$^{-2}$, respectively. This is consistent with the location of the GRB, well outside the galaxy's light.

**Origin of the X-ray afterglow**

*Swift* observations show a rapidly fading X-ray afterglow followed by a shallower decline $F_X \propto t^\alpha$ with $\alpha=1.11^{+0.08}_{-0.07}$ between 1 ks and 40 ks, and a final steep decay with $\alpha=3 \pm 0.5$ after 40 ks. Based on this model, we infer an X-ray flux of ≈4 × 10$^{-12}$ erg cm$^{-2}$ s$^{-1}$ at 11 hrs. This corresponds to a luminosity $L_X$≈6×10$^{43}$ erg s$^{-1}$ at 346 Mpc, nearly two orders of magnitude below the typical X-ray luminosity of cosmological GRB afterglows at this epoch (cf. Figure 7 of ref. 23). The low ratio between the observed X-ray flux and the emitted gamma-ray fluence, log $f_{X,11hr}/F_\gamma$ ~ -7.9, is indicative of atypical properties for this explosion (cf. Figure 17 of ref. 12).

Our SED analysis (Figure 2) demonstrates that the X-ray counterpart is dominated by non-thermal emission consistent with synchrotron radiation. While we interpret the early (<300 s) X-ray emission as the tail of the prompt phase, at later times (>1,000 s) the most common origin of non-thermal afterglow radiation is the interaction between the ambient medium and the GRB jet occurring at large distances (>$10^{17}$ cm) from the central source. In this external shock model[61], a flux decay rate of 2 or faster is explained by geometrical factors due to the collimation of the GRB outflow[62]. The time $t_j$ at which the light curve steepens, the so-called jet-break, increases with the jet opening angle $\theta_c$. A jet-break at 40 ks would require a very narrow jet, and then can only achieve a decay of $\alpha=p\approx2.1$, in mild tension with the observations. We tested the hypothesis of an early jet-break by modeling the X-ray and early (~$T_0$+1 hr) UVOT data with *afterglowpy*[63] assuming a uniform external environment and both a top-hat and a Gaussian lateral structure for the jet. Despite the dataset being limited, it provides tight constraints to the model: the flat UVOT SED at $T_0$+1 hr (Figure 2) requires the synchrotron peak to lie close to the optical range, constraining the value of $\nu_m$ and $F_{pk}$; the X-ray spectrum places the cooling frequency at $\nu_c$> 10 keV and provides a measurement of $p$~2.0-2.1, and the X-ray light curve constrains the jet opening angle $\theta_c$ and the viewing angle $\theta_v$. We performed Bayesian parameter estimation with *emcee*[64] and nine free parameters: $E_{K,iso}$, $\theta_c$, $\theta_v$, an outer jet truncation angle $\theta_w$, $n$, $p$, $\varepsilon_e$, $\varepsilon_B$, and the participation fraction $\xi_N$. The best fit has a reduced chi-squared $\chi_\nu^2$~1.8, fits with $\xi_N$ frozen at 1 found similar $\chi_\nu^2$ but required unphysical shock parameters $\varepsilon_e \approx \varepsilon_B \approx 1$. The parameter estimation reports a jet of energy $E_{K,iso}$ ~ (0.8-17) ×$10^{51}$ erg, width $\theta_c$ ~ 1.9-5.7 deg, viewed at $\theta_v$ ~ 1.1-5.4 deg from the jet axis. The external density is $n$ ~ 0.016 - 12 cm$^{-3}$. The shock parameters are $p$ ~ 2.1-2.2, $\varepsilon_e$ ~ 0.05 - 0.77, $\varepsilon_B$ ~ (0.1 - 6.0) ×$10^{-4}$, and $\xi_N$ ~ (0.8-9.6)×$10^{-2}$. The beaming-corrected kinetic energy of the jet in this scenario is 0.4 - 4.4 ×$10^{49}$ erg. Assuming the angular size-corrections between the afterglow and prompt emissions are similar, this scenario gives ~65% probability to an unphysical gamma-ray efficiency $\eta_\gamma = E_{\gamma,iso}/E_{K,iso}$ > 100% and a 90% probability $\eta_\gamma$ > 15%. In combination with the poor reduced chi-squared of 1.8 we conclude it is challenging for an external shock to simultaneously reproduce the salient features of the GRB afterglow: a flat UV/optical spectrum at $T_0+1$ hr, an X-ray spectrum $\beta_X \sim 0.5$, and a steep decay of the X-ray flux after 40 ks, while remaining within the energetic limits of the prompt emission. This tension may be alleviated when considering the effects of Inverse Compton (IC) cooling. In the limit of Thompson-scattering dominated IC cooling[65], we estimate that the required isotropic energy would increase by a factor of ~100, and the density decreased by a factor of ~1,000. However, the jet opening and viewing angles would have to decrease down to 0.5 deg to reproduce the final steep decay.

If not caused by a jet break, a rapid drop in brightness is difficult to produce due to the relativistic and extended nature of the GRB outflow. Due to the curvature effect[13] any rapid decrease in brightness in the lab-frame of the GRB will be smeared out in the observer frame due to different arrival times of the photons, producing a decay of $\alpha=2+\beta_X\sim2.5$. Nevertheless, this is a steeper slope than that allowed by the jet-break model and may present a better description than the standard external shock. If interpreted as a curvature effect, the steepening at 0.5 d links the observed X-ray emission either to long-lasting activity of the central engine, as in the "internal plateau" model[66,67], or to the angular structure of the GRB jet. If a structured jet produces GRB prompt emission in the high latitude regions (the jet "wings"), this emission would be Lorentz de-boosted relative to the core prompt emission and delayed via the curvature effect[36]. With appropriate jet structures, this can manifest as X-ray emission with a shallow decay followed by a steep declining light curve. This feature, normally hidden by the brighter external shock emission, may become apparent in the case of a "naked'' structured GRB exploding in a rarefied medium. This latter model offers a consistent explanation of the X-ray behavior of GRB211211A and its physical offset from the galaxy without the requirement of hours-long activity of the central engine.

Despite uncertainty in the physical origin of the afterglow emission, the observed X-ray spectrum is well measured and its extrapolation to the UVOIR bands unambiguously places it below the UV/optical detections after ~$T_0$+5 hr. The observed UVOIR excess was measured by subtracting this extrapolated non-thermal component. This procedure does not require a physical interpretation of the non-thermal emission and provides an upper bound on the non-thermal contribution in the UVOIR bands. Thus, the identification of the UVOIR excess does not depend on the specific physical interpretation of GRB211211A's non-thermal emission.

**Origin of the UVOIR excess**

**Collapsar model** - We first examine the most common case of a long GRB produced by the collapse of a rapidly rotating massive star (collapsar). The emergence of the SN blast wave can produce a luminous blue emission in excess of the standard afterglow[25], and we test whether this is consistent with the observed UVOIR excess in GRB211211A. Collapsars arise from compact stellar cores and produce energetic and long-lived Ic supernovae/hypernovae. However, if the collapsar engine does not produce significant $^{56}$Ni (e.g. from a fallback collapsar), the blastwave produces a short-lived SN light curve that dies out in the first 10 d. To test this model, we ran a series of hypernova explosions, varying the mass (2.5-40 $M_\odot$) and density profile (varying the slope in the density of the core and envelope) of the progenitor star as well as the explosion energy (spherically $10^{51}$-$10^{52}$ erg). Although we can reproduce the evolution of the bolometric luminosity (Extended Data Table 1), the early-time emission in our best-fit model is too energetic (in the UV and extreme UV). As the ejecta cools, the emission peaks in the IR at late times, but the luminosity is several orders of magnitude too dim to explain the observations. To account for the optical and IR emission, the photosphere of the rapidly expanding SN must uncover the collapsar accretion disk and wind ejecta from this disk must have similar-enough properties to NS merger disks[68,69] to produce a kilonova-like transient. However, even in this case, the large mass reservoir of a collapsar would power a long-lived late-peaking transient, not consistent with the observations. For the collapsar model to work, we must also explain the offset of the GRB from its host galaxy. O/B stars in binaries can be unbound during the supernova explosion of the primary star, imparting a "kick" of up to 200 km s$^{-1}$ onto the O/B companion[70]. This proper motion could move the companion O star well beyond its star forming region (~1 kpc in 5 Myr), but it is unlikely that this kick is sufficient to explain the large offset of this burst. In summary, a massive star progenitor for GRB211211A would naturally account for its long duration but requires a combination of unusual circumstances (a low $^{56}$Ni yield explosion, a low-mass neutron rich disk outflow, and an extreme kick velocity) to explain the entire set of observations.

**Compact binary merger model** - The observed excess emission is much better fit by the ejecta from a compact binary merger, composed either of two NSs or a NS and a stellar mass black hole (BH). Figure 3 shows the range of model predictions consistent with the observations: only a small subset of light curves (4 out of 900 in the "on-axis" angular bin; $\theta_v$~0-16 deg) match the observing constraints. The nIR luminosities are well described by dynamical ejecta of mass $M_d$~0.01-0.03 $M_\odot$, lower than the value inferred for GRB060614[7,8]. The bright UV/optical counterpart suggests a massive (>0.01 $M_\odot$) wind component to the kilonova ejecta. However, the time-dependent spectra from the Los Alamos National Laboratory (LANL) grid of kilonova models[28] produce light curves that are too dim to match the observed UV/optical luminosities or require too large an ejecta mass (~0.1 $M_\odot$). Models with large ejecta mass ($M_w$~0.1 $M_\odot$) fit better the early time data but overpredict the fluxes at later times and, vice versa, the model with lower ejecta mass ($M_w$~0.01 $M_\odot$) provides a good description of the dataset only after ~11 hrs. All consistent models adopt a toroidal morphology for the high-opacity ejecta and a polar outflow of low-opacity ejecta and high expansion velocity $v_w$~0.3 $c$.

It is likely that a number of alterations to the kilonova ejecta mechanism can help explain the early excess emission. For example, we have not conducted a detailed study varying the composition that changes both the opacity and the radioactive heating. Uncertainties in radioactive energy deposition[71] and in the properties of the disk-wind ejecta allow for a wide range of behaviors and our study here only touches the surface of all possibilities. However, in its simplest form, a radioactive-powered kilonova captures the late-time evolution of the observed UVOIR transient but has difficulties in reproducing the bright optical emission seen at early times ($T_0$+0.2 d).

An alternative way to alleviate the requirement on the ejecta mass is that the kilonova is powered by an additional energy source or affected by the jet-ejecta interactions[33]. To study the engine-powered models, we used the same method as in previous studies[31]. For central power sources, either a magnetar or fallback accretion on the central BH, the energy must transport out from the center to affect the light-curves. In these models[31], the central power sources do not alter the emission until ~5 d after the merger for wind mass ~0.01 $M_\odot$. However, if the jet is able to evacuate a region above the compact remnant, this delay can be reduced. We mimicked this evacuation by a series of spherically symmetric models, reducing the total wind mass to ~$10^{-7}$ $M_\odot$. Although the signal peaks earlier it is still too late to explain our observations and the resultant spectrum is too high energy (peaking in the extreme UV; Extended Data Figure 6). Turbulent motion may help to accelerate the UV peak by advecting the energy toward the outer layers more rapidly.

Although we caution that kilonova models are affected by large systematic uncertainties, we find that the majority of engine-driven kilonova models[31,72,73] peak several hours/days after the merger, whereas jet-ejecta interactions remain a plausible solution to enhance the early emission.

In summary, we find that a compact binary merger would naturally account for most of the observed features of GRB211211A, from the onset of its kilonova to its environment and high-energy properties. The main challenge to this model remains the long duration of the prompt gamma-ray emission, requiring an active central engine for up to ~100 s.

**Data Availability**
Data from NASA's missions are publicly available from the High Energy Astrophysics Science Archive Research Center (HEASARC) at https://heasarc.gsfc.nasa.gov. *Swift* XRT products are available from the online GRB repository https://www.swift.ac.uk/xrt_products. Other data are available from the corresponding author upon reasonable request. The broad grid of kilonova models is publicly available from the Center for Theoretical Astrophysics Los Alamos National Laboratory at https://ccsweb.lanl.gov/astro/transient/transients_astro.html

**Code Availability**
Results can be reproduced using standard free analysis packages. Methods are fully described. Code used to produce figures can be made available upon request.

**Additional References**

42. Arnaud, K. A. XSPEC: the first ten years. *Astron. Soc. Pac. Conf. Ser.* **101**, 17–20 (1996)
43. Norris, J. P. & Bonnell J. T. Short Gamma-Ray Bursts with Extended Emission. *Astrophys. J.* **643**, 266-275 (2006)
44. Golkhou, V. Z. & Butler N. R. Uncovering the Intrinsic Variability of Gamma-Ray Bursts. Astrophys. J. 787, 90 (2014)
45. Amati, L. et al., Intrinsic spectra and energetics of BeppoSAX Gamma-Ray Bursts with known redshifts. *Astron. Astrophys.* **390**, 81-89 (2002)
46. Kouveliotou, C. et al., Identification of Two Classes of Gamma-Ray Bursts. *Astrophys. J.* **413**, L101 (1993)
47. Bloom, J. S., Kulkarni, S. R. & Djorgovski, S. G. The Observed Offset Distribution of Gamma-Ray Bursts from Their Host Galaxies: A Robust Clue to the Nature of the Progenitors. *Astrophys. J.* **123**, 1111-1148 (2002)
48. Kewley, L. J., Nicholls, D. C. & Sutherland, R. S., Understanding Galaxy Evolution Through Emission Lines. Annu. Rev. Astron. Astrophys. 57, 511-570 (2019)
49. Palmerio, J. T. et al., Are long gamma-ray bursts biased tracers of star formation? Clues from the host galaxies of the Swift/BAT6 complete sample of bright LGRBs. III. Stellar masses, star formation rates, and metallicities at z > 1. *Astron. Astrophys.* **623**, A26 (2019)
50. Freedman, W. L. et al., Final Results from the Hubble Space Telescope Key Project to Measure the Hubble Constant. *Astrophys. J.* **553**, 47-72 (2001)
51. Ilbert, O. et al., The VIMOS-VLT Deep survey - evolution of the galaxy luminosity function up to z = 2 in first epoch data. *Astron. Astrophys.* **439**, 863-876 (2005)

52. Peng, C. Y., Ho, L. C., Impey, C. D., & Rix, H.-W. Detailed Decomposition of Galaxy Images. II. Beyond Axisymmetric Models. *Astrophys. J.* **139**, 2097-2129 (2010)
53. Johnson, B. D., Leja, J., Conroy, C. & Speagle, J. S. Stellar Population Inference with Prospector. *Astrophys. J.* **254**, 22 (2021)
54. O'Connor, B. et al., A tale of two mergers: constraints on kilonova detection in two short GRBs at z ~ 0.5. *Mon. Not. R. Astron. Soc.* **502**, 1279-1298 (2021)
55. Whitaker, K. E., van Dokkum, P. G., Brammer, G. & Franx, M. et al., The Star Formation Mass Sequence Out to z = 2.5. *Astrophys. J.* **754**, L29 (2012)
56. Izzo, L. et al., The MUSE view of the host galaxy of GRB 100316D. *Mon. Not. R. Astron.* **472**, 4480-4496 (2017)
57. Wiersema, K. et al., The nature of the dwarf star-forming galaxy associated with GRB 060218/SN 2006aj. *Astron. Astrophys.* **464**, 529-539 (2007)
58. Leibler, C. N. & Berger E. The Stellar Ages and Masses of Short Gamma-ray Burst Host Galaxies: Investigating the Progenitor Delay Time Distribution and the Role of Mass and Star Formation in the Short Gamma-ray Burst Rate. *Astrophys. J.* **725**, 1202-1214 (2010)
59. Roman, M. et al., Dependence of Type Ia supernova luminosities on their local environment. *Astron. Astrophys.* **615**, A68 (2018)
60. Shen, R.-F. & Zhang, B. Prompt optical emission and synchrotron self-absorption constraints on emission site of GRBs. *Mon. Not. R. Astron. Soc.* **398**, 1936-1950 (2009)
61. Rees, M. J. & Meszaros, P. Relativistic fireballs - Energy conversion and time-scales.. *Mon. Not. R. Astron. Soc.* **258**, 41 (1992)
62. Rhoads, J. E. The Dynamics and Light Curves of Beamed Gamma-Ray Burst Afterglows. *Astrophys. J.* **525**, 737-749 (1999)
63. Ryan, G., van Eerten, H., Piro, L. & Troja, E. Gamma-Ray Burst Afterglows in the Multimessenger Era: Numerical Models and Closure Relations. *Astrophys. J.* **896**, 166 (2020)
64. Foreman-Mackey, D., Hogg, D. W., Lang, D. & Goodman, J. emcee: the MCMC hammer. *Publ. Astron. Soc. Pacif.* **125**, 306–312 (2013).
65. Sari, R. & Esin, A. A. On the Synchrotron Self-Compton Emission from Relativistic Shocks and Its Implications for Gamma-Ray Burst Afterglows. *Astrophys. J.* **548**, 787-799 (2001)
66. Zhang, B. & Mészáros, P. Gamma-Ray Burst Afterglow with Continuous Energy Injection: Signature of a Highly Magnetized Millisecond Pulsar. *Astrophys. J.* **552**, L35-L38 (2001)
67. Troja, E. et al. Swift Observations of GRB 070110: An Extraordinary X-Ray Afterglow Powered by the Central Engine. *Astrophys. J.* **665**, 599-607 (2007)
68. Siegel, D. M., Barnes, J. & Metzger, B. D. Collapsars as a major source of r-process elements. *Nature* **569**, 241-244 (2019)
69. Miller, J. M. et al. Full Transport General Relativistic Radiation Magnetohydrodynamics for Nucleosynthesis in Collapsars. *Astrophys. J.* **902**, 66 (2020)
70. Fryer, C., Burrows, A. & Benz W. Population Syntheses for Neutron Star Systems with Intrinsic Kicks. *Astrophys. J.* **496**, 333-351 (1998)
71. Barnes, J. et al. Kilonovae Across the Nuclear Physics Landscape: Impact of Nuclear Physics Uncertainties on r-process-powered Emission. *Astrophys. J.* **918**, 27 (2021)
72. Yu, Y.-W., Zhang, B. & Gao, H. Bright "Merger-nova" from the Remnant of a Neutron Star Binary Merger: A Signature of a Newly Born, Massive, Millisecond Magnetar. *Astrophys. J.* 776, L40 (2013)

73. Sarin, N., Omand C. M. B., Margalit B., Jones D. I. et al. On the diversity of magnetar-driven kilonovae Preprint at https://arxiv.org/abs/2205.14159 (2022)
74. Galametz, A. et al. CANDELS Multiwavelength Catalogs: Source Identification and Photometry in the CANDELS UKIDSS Ultra-deep Survey Field. *Astrophys. J., Suppl. Ser.* **206**, 10 (2013)
75. Pian, E. et al. An optical supernova associated with the X-ray flash XRF 060218. *Nature* **442**, 1011-1013 (2006)
76. Fremling, C. et al. The Zwicky Transient Facility Bright Transient Survey. I. Spectroscopic Classification and the Redshift Completeness of Local Galaxy Catalogs. *Astrophys. J.* **895**, 19 (2020)

**Acknowledgements**
This work was supported by the European Research Council through the Consolidator grant BHianca (Grant agreement ID: 101002761) and by the National Science Foundation (under award number 12850). BO was partially supported by the National Aeronautics and Space Administration through grants NNX16AB66G, NNX17AB18G, and 80NSSC20K0389. Part of this work was financially supported by Grants-in-Aid for Scientific Research 17H06362 (N.K.) from the Ministry of Education, Culture, Sports, Science and Technology (MEXT) of Japan. This work was partially supported by the Optical and Near-Infrared Astronomy InterUniversity Cooperation Program of the MEXT of Japan, and the joint research program of the Institute for Cosmic Ray Research (ICRR), the University of Tokyo. Research at Perimeter Institute is supported in part by the Government of Canada through the Department of Innovation, Science and Economic Development and by the Province of Ontario through the Ministry of Colleges and Universities. The development of afterglow models used in this work was partially supported by the European Union Horizon 2020 Programme under the AHEAD2020 project (grant agreement number 871158).

This work includes observations obtained at the international Gemini Observatory (PI: B. O'Connor; GS-2022A-Q-141), a programme of NSF's NOIRLab, which is managed by the Association of Universities for Research in Astronomy (AURA) under a cooperative agreement with the NSF on behalf of the Gemini Observatory partnership: the NSF (USA), National Research Council (Canada), Agencia Nacional de Investigación y Desarrollo (Chile), Ministerio de Ciencia, Tecnología e Innovacíon (Argentina), Ministério da Ciência, Tecnologia, Inovações e Comunicações (Brazil), and Korea Astronomy and Space Science Institute (Republic of Korea). The *HST* data (ObsID: 16846; PI: E. Troja) used in this work was obtained from the Mikulski Archive for Space Telescopes (MAST). STScI is operated by the Association of Universities for Research in Astronomy, Inc., under NASA contract NAS5-26555. This research is partially based on observations under our ToO proposal number DOT-2021-C2-P71 (PI: R. Gupta) and proposal number DOT-2021-C2-P54 (PI: S. B. Pandey) obtained at the 3.6m Devasthal Optical Telescope (DOT), which is a National Facility run and managed by Aryabhatta Research Institute of Observational Sciences (ARIES), an autonomous Institute under Department of Science and Technology, Government of India. RG and SBP thankfully acknowledges all the observing and support staff of the 3.6m DOT and 1.3m DFOT to maintain and run the observational facilities at Devasthal Nainital. RG is especially thankful to Dr. B. Kumar, Amit Kumar Ror and M. Sarkar for the observations with the 3.6m DOT and 1.3m DFOT under the approved observing proposals: DOT-2021-C2-P71 (PI: R. Gupta), DFOT-2021B-P29 (PI: R. Gupta), and DOT-2021-C2-P54 (PI: S. B. Pandey). RG, AA, KM, and SBP acknowledge the BRICS grant (DST/IMRCD/BRICS

/PilotCall1/ProFCheap/2017(G)) for the financial support. RG and SBP also acknowledge the financial support of ISRO under the AstroSat archival Data utilization program (DS_2B-13013(2)/1/2021-Sec.2). A.A. acknowledges funds and assistance provided by the Council of Scientific & Industrial Research (CSIR), India with file no. 09/948(0003)/2020-EMR-I. MCG acknowledges support from the Ramón y Cajal Fellowship RYC2019-026465-I. YDH acknowledges support under the additional funding from the RYC2019-026465-I. AJCT acknowledges support from the Spanish Ministry Project PID2020-118491GB-I00, Junta de Andalucia Project P20_01068 and the "Center of Excellence Severo Ochoa" award for the Instituto de Astrofísica de Andalucía (SEV-2017-0709). Based on observations collected at the Centro Astronómico Hispano-Alemán (CAHA) at Calar Alto (ToO Program 21B-2.2-028: PI: Castro-Tirado), operated jointly by Junta de Andalucía and Consejo Superior de Investigaciones Científicas (IAA-CSIC). These results made use of the Lowell Discovery Telescope (LDT) at Lowell Observatory. Lowell is a private, non-profit institution dedicated to astrophysical research and public appreciation of astronomy and operates the LDT in partnership with Boston University, the University of Maryland, the University of Toledo, Northern Arizona University and Yale University. Figure 1 was created with the help of the NOIRLab/IPAC/ESA/STScI/CfA FITS Liberator. We made use of IRAF, which is distributed by the NSF NOIRLab. This work made use of data supplied by the UK *Swift* Science Data Centre at the University of Leicester.

**Author Contributions**
ET initiated the project, coordinated the observations, their physical interpretation, and was the primary author of the manuscript. BO led the study of the GRB environment, contributed to afterglow modeling, physical interpretation and manuscript writing. CF and RW developed the supernova and kilonova models for this paper, contributing to physical interpretation and manuscript writing. GR performed the afterglow fits and contributed to manuscript writing. SD led the spectral analysis of the prompt gamma-ray emission and contributed to manuscript writing, JN calculated the temporal lags, NB derived the minimum variability timescale. AK, RG, AA, KM, SBP acquired and reduced the data of the 3.6m DOT telescope, RG and SBP contributed to the study of the prompt emission. NI, NK, RH, KLM, MN acquired and reduced the data of the MITSUME telescope. HJvE contributed to afterglow modeling and physical interpretation. EAC compared the kilonova models to the data. YDH, MCG, AJCT acquired and reduced the data of the CAHA telescope. All authors contributed to edits to the manuscript.

**Competing Interests**
The authors declare that they have no competing financial interests.

**Additional Information**
Supplementary Information is available for this paper.

Correspondence and requests for materials should be addressed to eleonora.troja@uniroma2.it and oconnorb@gwmail.gwu.edu

Reprints and permissions information is available at www.nature.com/reprints.

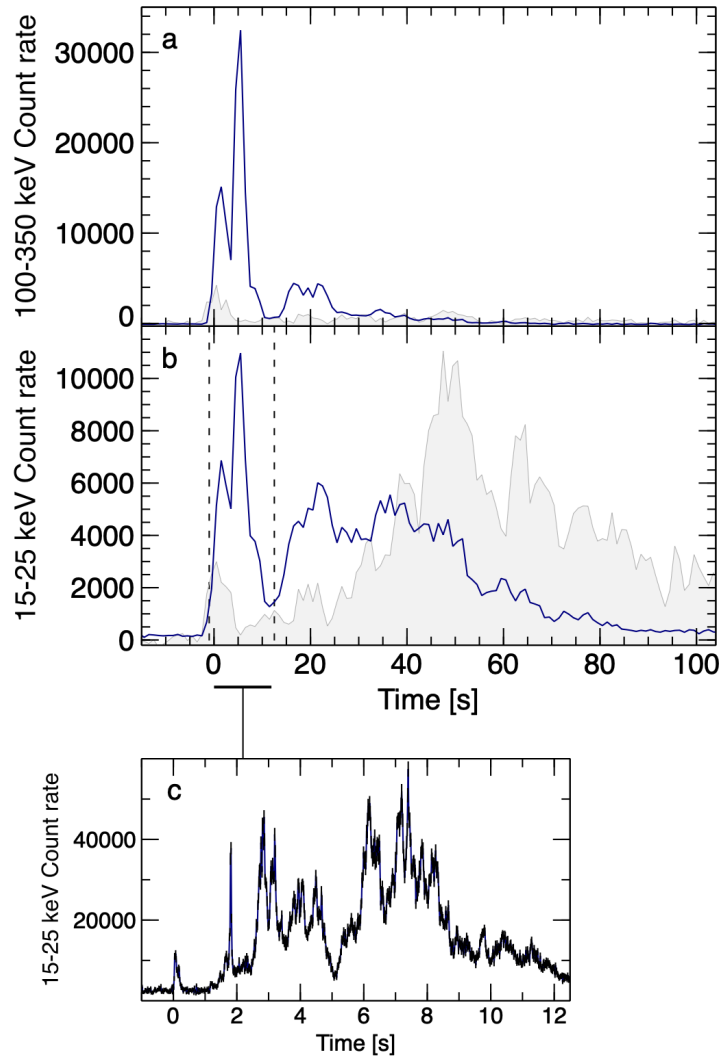

**Extended Data Figure 1 - Prompt gamma-ray phase of GRB211211A**

The *Swift* background subtracted light curves of GRB211211A are shown in two energy bands and compared with the time history of GRB060614 (gray shaded area) rescaled at a distance of 346 Mpc. The time bin is 1 s. Error bars are 1$\sigma$. Both bursts display a first episode with hard spectrum (dominant in GRB211211A), followed by a long-lasting tail with soft spectrum (dominant in

GRB060614). The inset zooms in the first 12 s, showing a weak precursor at $T_0$ preceding the main prompt event. The time bin is 16 ms.

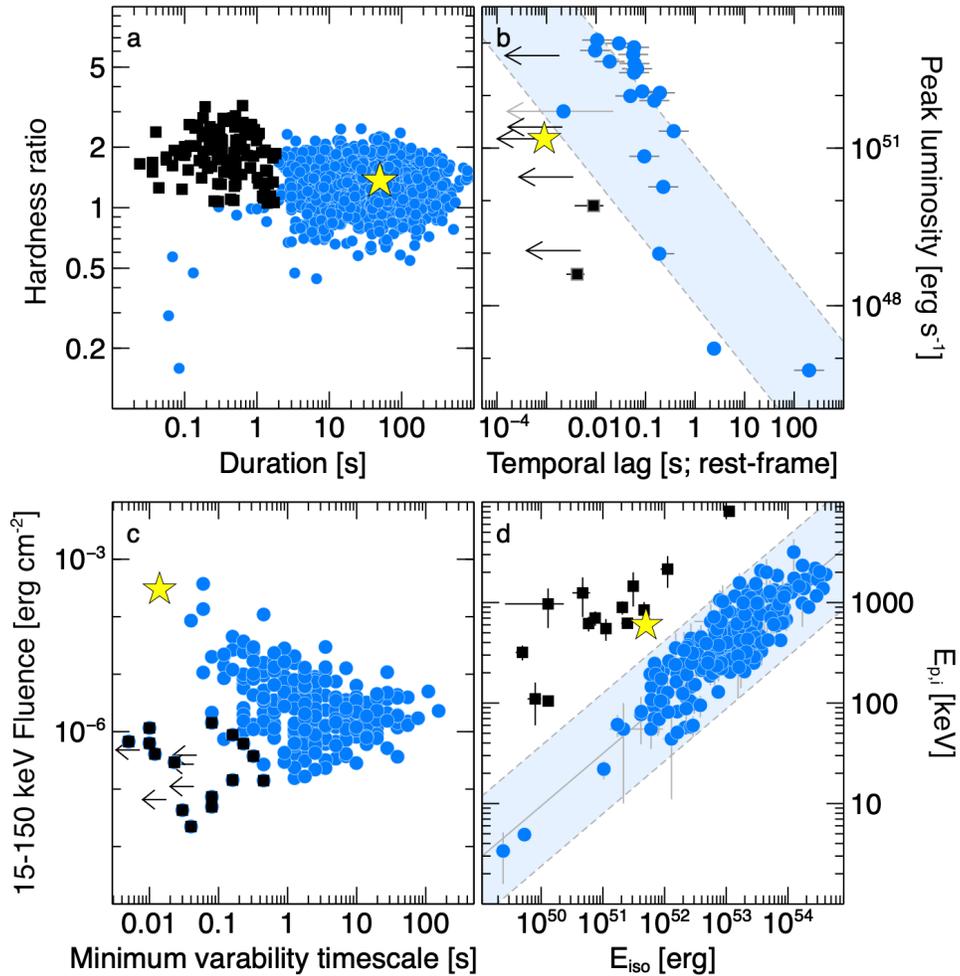

**Extended Data Figure 2 - GRB classification scheme**

The traditional GRB classification, based on the duration/hardness ratio diagram (a), is not unambiguous. Additional classificators, used to break the degeneracy, are the lag-luminosity relation (b), the variability timescale (c), and the Amati relation (d). Long GRBs (circles) and short GRBs (squares) occupy different regions of these plots. Dashed lines show the boundaries of the long GRB regions (shaded areas). GRB211211A (star symbol) belongs to the class of long soft bursts (a), but its other high-energy properties are common among short GRBs. Error bars represent 1σ; upper limits (arrows) are 3σ.

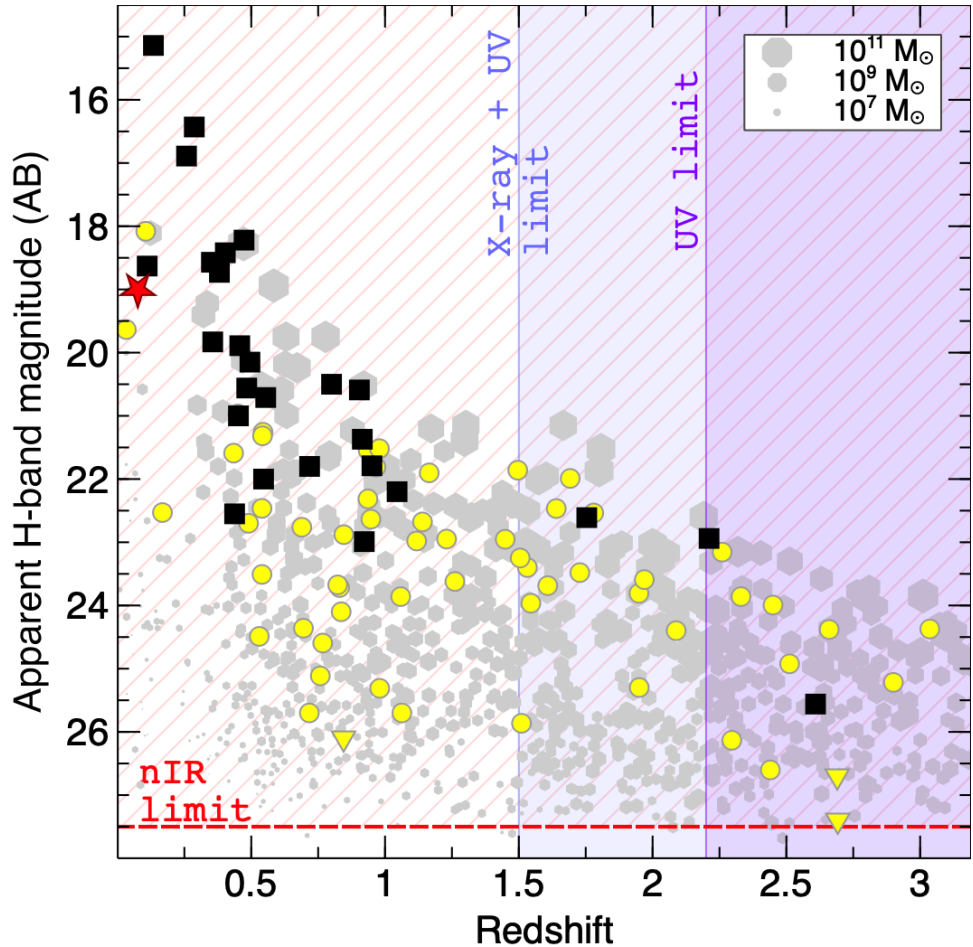

**Extended Data Figure 3 - Constraints on the distance scale of GRB211211A**

The nIR brightness of GRB host galaxies (short GRBs: squares; long GRBs: circles) is reported as a function of their redshift. For comparison, a randomly selected sample of field galaxies from the CANDELS survey[74] is shown in the background (octagons, with the symbol size proportional to the galaxy mass). The non-detection of an underlying galaxy in deep *HST F160W* imaging rules out most of the parameter space occupied by GRB hosts (hatched area). Additional constraints from the UV and X-ray afterglow rule out the case of a GRB in a distant ($z>1.5$; shaded areas) faint galaxy. These observations support the physical association between GRB211211A and the nearby galaxy at $z=0.0762$ (star symbol). Upper limits (downward triangles) are $3\sigma$.

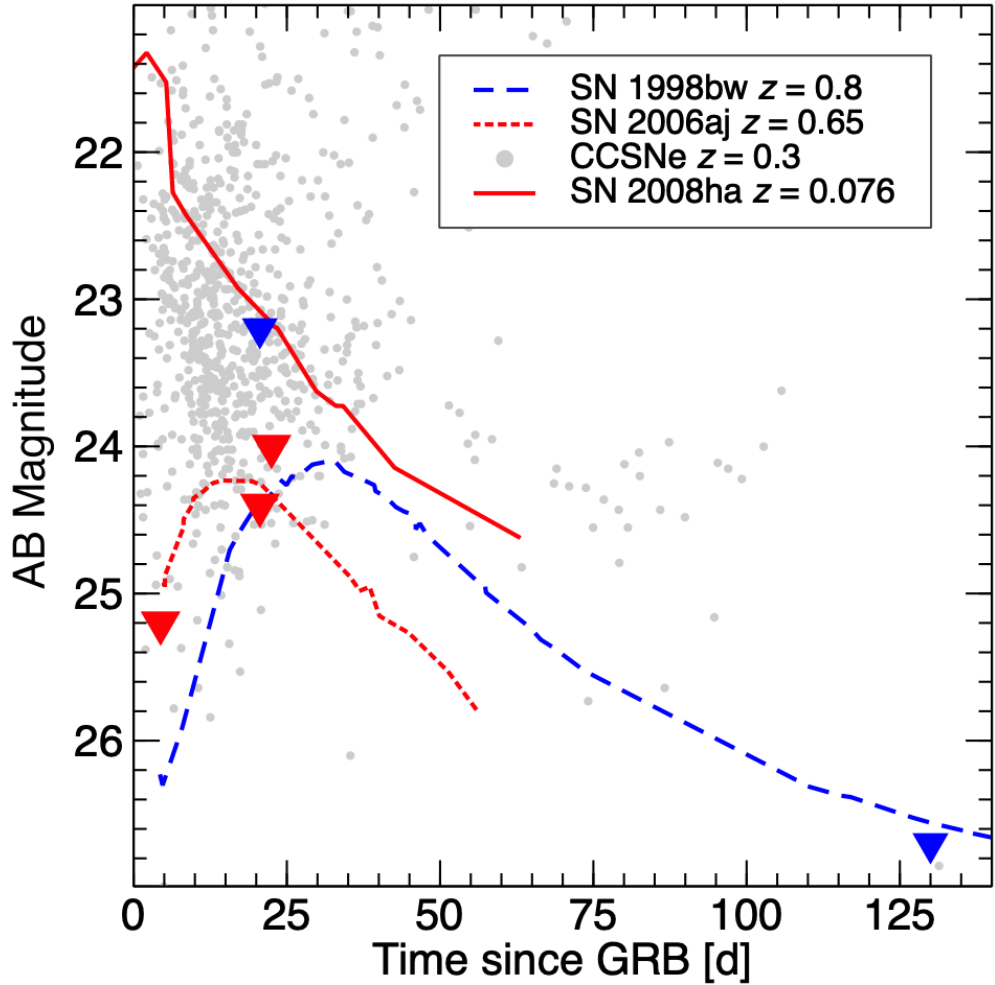

**Extended Data Figure 4 - No supernova associated with GRB211211A**

Optical upper limits (representing the $3\sigma$ confidence level) in the r-band (red) and i-band (blue) rule out the presence of any known supernova following GRB211211A. Bright SNe associated with GRBs, such as SN1998bw[15] and SN2006aj[75], would have been detected up to $z=0.8$ (dashed line) and $z=0.65$ (dotted line), respectively. Symbols show the peak magnitude of core-collapse SNe from the ZTF Bright Transient Survey Sample[76] rescaled at $z=0.3$, demonstrating that most ordinary SNe were detectable up to this distance. At the distance $z\sim0.076$ of the putative host galaxy, the faint SN2008ha[6] (solid line) is also ruled out.

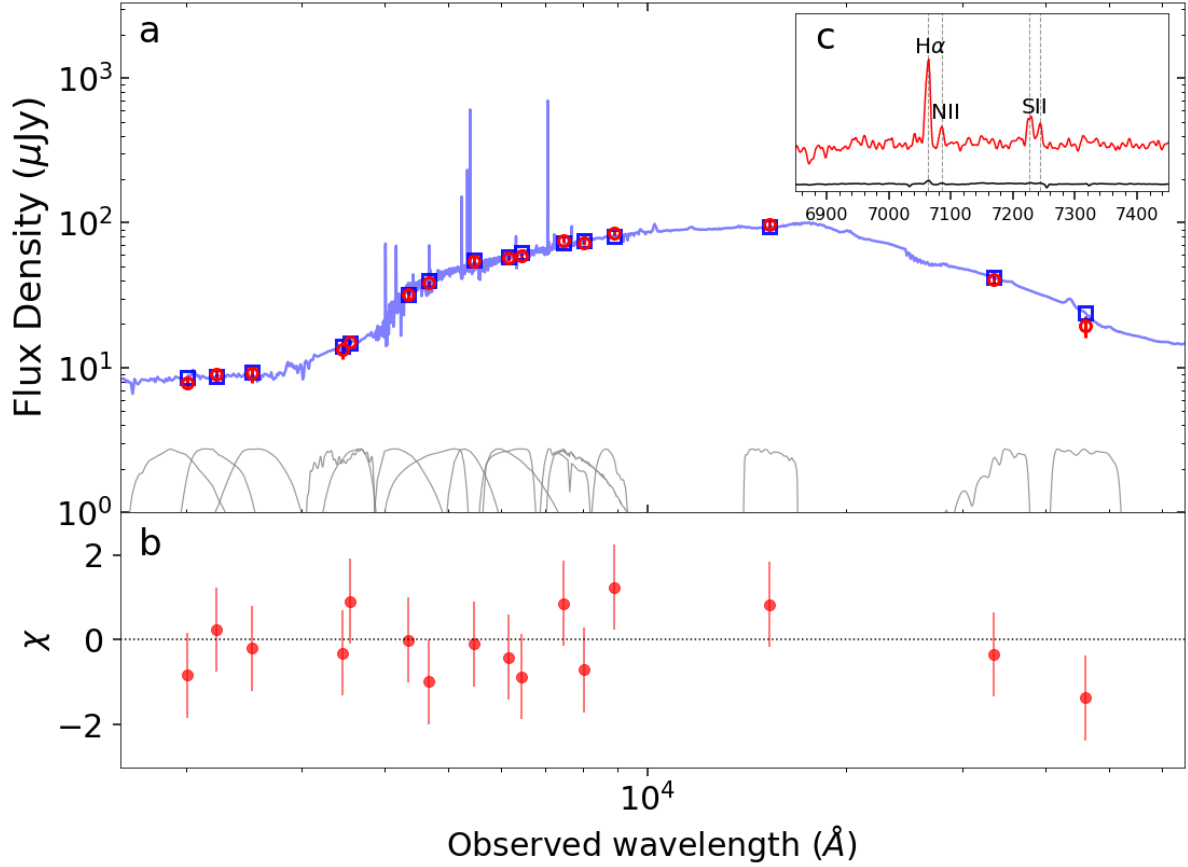

**Extended Data Figure 5 - Host galaxy spectral energy distribution**

(a) The model SED (blue line) and model photometry (blue squares) derived using Prospector is compared to the observed photometry (red circles). The fit residuals (b) are displayed in the bottom panel. The inset (c) displays a Gemini/GMOS-S spectrum of the host galaxy in the vicinity of Hα, [NII], and the [SII] doublet, yielding $z\sim0.0762$. Error bars represent $1\sigma$.

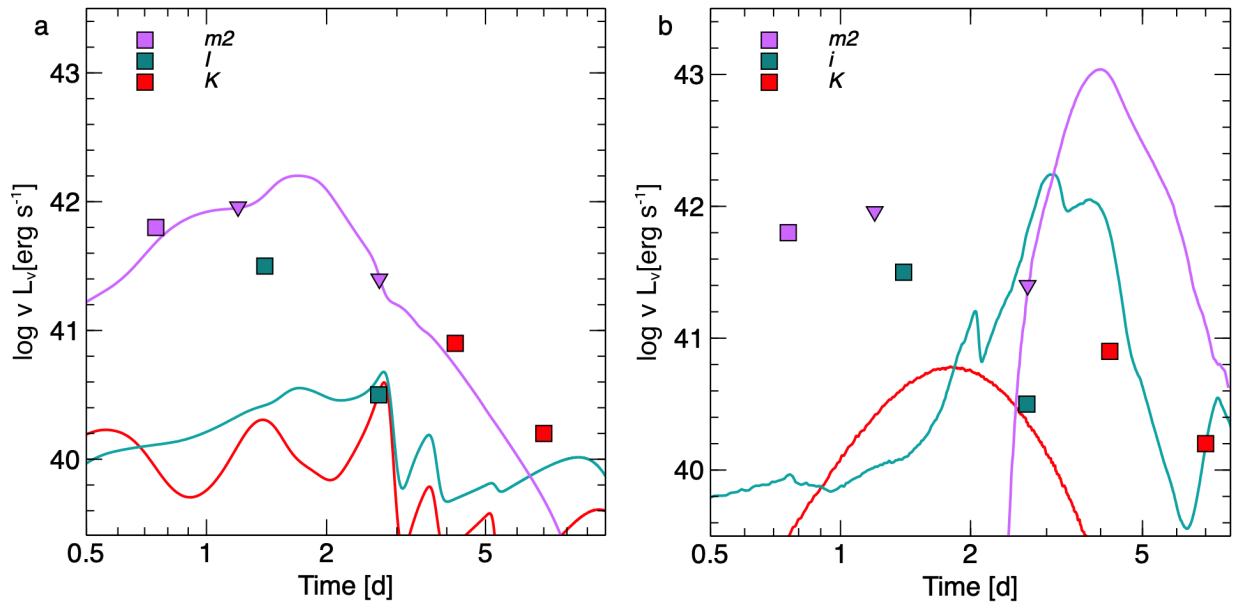

**Extended Data Figure 6 - A comparison of models for luminous blue transients**

The UVOIR counterpart is compared with a set of models producing luminous ($L_{bol} \sim 10^{42}$ erg s$^{-1}$) and short-lived (< 7 d) transients: (a) a low-nickel SN from a fallback collapsar[25] underpredicts the optical/nIR emission; (b) a magnetar-powered kilonova[31] does not easily reproduce the timescales and colors.

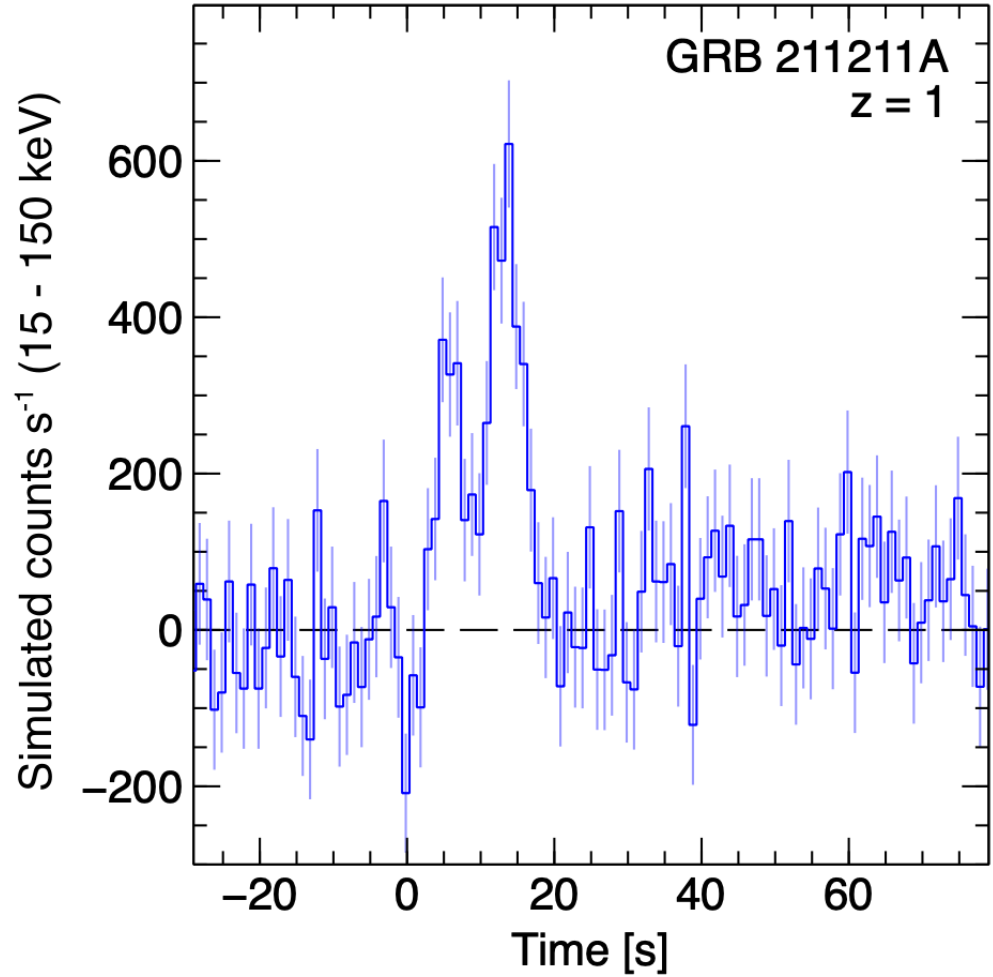

**Extended Data Figure 7 - GRB211211A at high redshift**

Simulated *Swift* lightcurve in the 15-150 keV energy band (observer's frame) of GRB211211A assuming $z=1$, an incident angle of 45 deg, and ~6,000 cts s$^{-1}$ background rate. Time bin is 1 s, error bars represent 1$\sigma$. An event similar to GRB211211A would be detected up to redshift $z$~1 and beyond. At these distances, it would appear as a standard long GRB with a duration of ~20 s.

**Extended Data Table 1 – Best-fit black body parameters**

Temporal evolution of the kilonova properties estimated from blackbody fitting.

| T - $T_0$ (d) | Bolometric Luminosity ($10^{42}$ erg s$^{-1}$) | Temperature ($10^3$ K) | Radius ($10^{15}$ cm) |
|---|---|---|---|
| 0.2 | 3.5 ± 2.0 | 16 ± 5 | 0.28 ± 0.14 |
| 0.4 | 1.90 ± 0.15 | 8.0 ± 0.3 | 0.80 ± 0.05 |
| 1.4 | 0.37 ± 0.10 | 4.9 ± 0.5 | 0.9 ± 0.2 |
| 4.2 | 0.13 ± 0.02 | 2.50 ± 0.10 | 2.0 ± 0.2 |

## Supplementary Methods

**Observations and Data Reduction**

**Swift** - The *Swift* X-ray Telescope (XRT) began observations of GRB211211A at 69 s after the GRB trigger and continued monitoring for ~3 days. The X-ray lightcurve and spectra were retrieved from the *Swift* XRT GRB repository[77]. The GRB counterpart was detected by the UltraViolet and Optical Telescope (UVOT) in all filters. UVOT data were analyzed using HEASOFT v.6.30. Individual exposures from the same epoch were aligned and then combined using the task *uvotimsum*. To minimize the contamination from the nearby galaxy G1, we determined the source count-rates in 3" circular apertures with the background estimated in a concentric source-free annular region. We used an aperture correction to determine the equivalent source count-rates in a 5" aperture, and then applied standard UVOT magnitude zeropoints[78]. Late-time images were used to determine the host galaxy brightness using a 5" circular aperture. The photometry of the transient and its host galaxy is reported in Extended Data Table 1.

**Optical/nIR imaging** - Our observations of GRB211211A began at $T_0+5.1$ hrs with the 50-cm Multicolor Imaging Telescopes for Survey and Monstrous Explosions (MITSUME[79,80,81]) acquiring 110×60 s simultaneous images in $g$, $R_c$, and $I_c$ filters. Observations were carried out for three consecutive nights, but the GRB counterpart was only detected on the first night. The data were reduced using the MITSUME pipeline[82] and image subtraction between the first, second and third epoch was performed using HOTPANTS[83]. Upper limits were derived by planting artificial sources of known brightness, then repeating the image subtraction step.

Starting on 2021 Dec 11 UT 22:11:29 (~$T_0$ + 9.0 hr), multi-band (*U*, *B*, *V*, *R*, *I*) optical observations of GRB211211A were carried out using the 4Kx4K CCD Imager mounted at the axial port of the 3.6m Devasthal Optical Telescope (DOT[84,85,86,87]). Observations continued until twilight time (~$T_0$ + 11.4 hr) and were repeated in subsequent nights until the optical counterpart was no longer detected. Starting on 12 Dec 2021 at 3:53 UT ($T_0$ + 15 hr), additional imaging of the GRB field was carried out with the 2.2m telescope, equipped with the Calar Alto Faint Object Spectrograph (CAFOS), at the Calar Alto Observatory in Almería (Spain) using the *g'r'i'* Sloan filters.

Additional epochs of observations were performed between 2021 Dec 31 and 2022 Jan 02 to search for the possible SN peak. We acquired multiple *R*-band images (24×300s) with the 1.3m Devasthal Fast Optical Telescope (DFOT) and deep *g'r'i'* exposures with the CAHA telescope. At late times, we targeted the field with the Large Monolithic Imager (LMI) on the 4.3-m Lowell Discovery Telescope (LDT) using the *u*, *g*, *r*, *i*, and *z* filters in order to refine the photometry of its host galaxy.

The data were pre-processed using standard CCD reduction techniques including bias subtraction, flat-fielding, fringe correction and cosmic ray rejection. Aperture photometry on the GRB counterpart was performed using Source Extractor[88] and selecting circular apertures with radius 1.5 times larger than the image full width half maximum. Forced photometry was performed on images acquired at later times (>$T_0$ + 2 d). Host galaxy magnitudes were derived using Kron-like elliptical apertures (MAG AUTO). The photometric zeropoints were calibrated using the same nearby point sources in the Sloan Digital Sky Survey (SDSS[89]) and transformed using empirical equations[90].

We imaged the field with the *Hubble Space Telescope* (*HST*) using the WFC3/UVIS and IR cameras with the *F814W* and *F160W* filters, respectively. Observations were carried out between April 2 and April 21, 2022 (~$T_0$ + 4 months). The data were reduced using standard procedures within the DrizzlePac software[91] in order to align, drizzle, and combine exposures. We used Source Extractor to detect sources and perform aperture photometry. For each filter, we used the zeropoints stored in the keyword PHOTFLAM.

**Gemini Spectroscopy** - We utilized the Gemini Multi-Object Spectrograph (GMOS) mounted on the 8.1-m Gemini South telescope to obtain a series of 4×600 s spectra, using the R400 grating and a 1" slit. We chose a slit position angle (PA) of 64 deg such that the slit covers both the host galaxy's center and the GRB's optical position. The data were reduced and analyzed using standard procedures in Gemini IRAF. In order to correct for slit losses, the flux-calibrated spectra were matched to the photometry of the host galaxy. Line fluxes were derived by fitting each line with a Gaussian function and estimating the continuum from nearby spectral regions.

**White dwarf merger model**

One of the possible progenitors adopted to explain SN-less long GRBs is the merger of a white dwarf (WD) with either a NS or a stellar mass BH[92,93]. These are old stellar systems, and their mergers produce accretion disks with longer accretion timescales than those from compact binary mergers[94]. Therefore, they can explain the lack of SN, the long gamma-ray duration, and the environment of hybrid GRBs. The merger ejecta contains a moderate amount (< 0.1 $M_{sun}$) of radioactive $^{56}$Ni powering fast-evolving (weeks to month long) optical transients[95] with luminosities in the range $10^{40}$ - $10^{43}$ erg s$^{-1}$. However, past light curve calculations[96] of these electromagnetic counterparts resemble faint type Iax SNe, and do not match the colors and timescales of the excess emission in GRB211211A.

**Rate of events**

In 17 years of *Swift* operations at least 2 hybrid GRBs were identified, GRB211211A and GRB060614. This allows us to place a lower limit to the rate of hybrid GRBs:

$$R > 0.7 \frac{\Omega}{4\pi} \frac{1}{V_z} \frac{1}{\varepsilon T} \frac{1}{\eta} \approx 0.04 \, Gpc^{-3} \, yr^{-1} \quad (1)$$

where $\Omega \sim 2.2$ sr is the *Swift* field of view for partial coding >10%, $T \sim 17$ yr the mission lifetime, $\varepsilon \sim 78\%$ its duty cycle[11], $\eta \sim 17/20$ the efficiency of SN searches, and $V_z \sim 7.3$ Gpc$^3$ the comoving volume within $z\sim0.3$. The factor 0.7 is the 68% c.l. lower limit on the number count derived from Poissonian statistics[97].

In a similar way we derive an upper limit to the rate by assuming that all the SN-less long GRBs within $z\sim0.3$ belong to the class of nearby hybrid bursts. We consider events with $E_{\gamma,iso} > 10^{49}$ erg in order to minimize selection effects due to the trigger efficiency. This only excludes one burst (GRB111005A). The 68% c.l. upper limit on the number count is ~12, from which we derive $R < 0.8$ Gpc$^{-3}$ yr$^{-1}$ using Supplementary Equation (1).

For comparison, the observed rate of short GRBs[40] ranges between 2.2 Gpc$^{-3}$ yr$^{-1}$ and 6.4 Gpc$^{-3}$ yr$^{-1}$ (68% c.l.). This value was derived for luminosities $L_{iso} > 5 \times 10^{49}$ erg s$^{-1}$, and therefore does not include the contribution of under-luminous off-axis bursts. The observed ratio of hybrid to short GRBs is simply given by the ratio of the two distributions (0.8%-26% at the 68% c.l.). However, the local population of GW counterparts is likely dominated by faint events seen off-axis[98], whose rate of detection depends on the distribution of jet opening angles and their angular profiles. We parameterize these properties using the beaming factor $f_b$, and caution that its value may differ between the population of bursts (hybrid and short), thus affecting their relative ratio in the nearby Universe.

# Supplementary Table

**Supplementary Table 1 – UV, optical and nIR observations of GRB211211A.**

Upper limits (u.l.) are 3 σ. Values are corrected for Galactic extinction in the direction of the burst, $A_V = 0.047$ mag[99].

| T - T$_0$ (d) | Exposure (s) | Telescope | Filter | Magnitude AB | Error (68% c.l.) |
|---|---|---|---|---|---|
| 0.0009 | 9 | UVOT | v | 17.3 | u.l. |
| 0.0012 | 36 | UVOT | White | 20.33 | 0.25 |
| 0.0020 | 96 | UVOT | White | 21.2 | u.l. |
| 0.042 | 195 | UVOT | u | 19.75 | 0.13 |
| 0.044 | 193 | UVOT | b | 19.62 | 0.2 |
| 0.046 | 185 | UVOT | White | 19.6 | 0.06 |
| 0.049 | 196 | UVOT | w2 | 19.59 | 0.12 |
| 0.051 | 195 | UVOT | v | 19.27 | 0.28 |
| 0.053 | 196 | UVOT | m2 | 19.58 | 0.18 |
| 0.056 | 196 | UVOT | w1 | 19.44 | 0.12 |
| 0.058 | 78 | UVOT | u | 19.38 | 0.16 |
| 0.19 | 877 | UVOT | u | 19.75 | 0.07 |
| 0.20 | 570 | UVOT | b | 19.79 | 0.17 |
| 0.24 | 3300 | MITSUME | g | 19.85 | 0.15 |
| 0.24 | 3300 | MITSUME | R$_c$ | 20.3 | 0.19 |
| 0.24 | 3300 | MITSUME | I$_c$ | 20.2 | u.l. |
| 0.25 | 467 | UVOT | m2 | 20.5 | 0.16 |
| 0.29 | 3300 | MITSUME | g | 20.19 | 0.16 |
| 0.29 | 3300 | MITSUME | R$_c$ | 19.99 | 0.18 |
| 0.29 | 3300 | MITSUME | I$_c$ | 19.98 | 0.19 |
| 0.37 | 200 | DOT | R | 20.17 | 0.06 |
| 0.38 | 300 | DOT | I | 20.26 | 0.08 |
| 0.38 | 300 | DOT | R | 20.03 | 0.05 |
| 0.38 | 300 | DOT | V | 20.18 | 0.07 |
| 0.39 | 300 | DOT | B | 20.51 | 0.07 |
| 0.39 | 360 | DOT | U | 20.78 | 0.07 |
| 0.40 | 300 | DOT | I | 20.17 | 0.08 |
| 0.40 | 300 | DOT | R | 20.14 | 0.04 |
| 0.40 | 300 | DOT | V | 20.26 | 0.04 |
| 0.41 | 300 | DOT | B | 20.45 | 0.06 |
| 0.41 | 360 | DOT | U | 20.75 | 0.07 |
| 0.42 | 200 | DOT | I | 20.2 | 0.09 |
| 0.42 | 200 | DOT | R | 20.15 | 0.05 |
| 0.42 | 200 | DOT | V | 20.28 | 0.04 |
| 0.42 | 200 | DOT | B | 20.49 | 0.05 |
| 0.43 | 360 | DOT | U | 20.76 | 0.07 |
| 0.43 | 200 | DOT | I | 20.08 | 0.08 |
| 0.43 | 200 | DOT | R | 20.24 | 0.04 |
| 0.44 | 360 | DOT | U | 20.81 | 0.07 |
| 0.45 | 200 | DOT | I | 20.17 | 0.09 |
| 0.45 | 200 | DOT | R | 20.29 | 0.05 |
| 0.45 | 200 | DOT | V | 20.35 | 0.05 |
| 0.46 | 200 | DOT | B | 20.53 | 0.05 |
| 0.46 | 360 | DOT | U | 20.97 | 0.07 |
| 0.46 | 200 | DOT | I | 20.34 | 0.09 |
| 0.47 | 200 | DOT | R | 20.41 | 0.09 |
| 0.47 | 200 | DOT | V | 20.27 | 0.07 |
| 0.47 | 200 | DOT | B | 20.68 | 0.15 |
| 0.62 | 900 | CAHA | i | 20.71 | 0.09 |
| 0.63 | 900 | CAHA | r' | 20.73 | 0.09 |
| 0.64 | 720 | CAHA | g' | 21.16 | 0.08 |
| 0.75 | 1771 | UVOT | w1 | 21.96 | 0.19 |
| 0.81 | 1771 | UVOT | w2 | 22.32 | 0.20 |
| 0.83 | 370 | UVOT | v | 19.8 | u.l. |
| 0.93 | 1247 | UVOT | u | 22.15 | u.l. |
| 0.98 | 1949 | UVOT | b | 22.3 | 0.4 |
| 1.22 | 1422 | UVOT | w1 | 22.3 | u.l. |
| 1.26 | 813 | UVOT | m2 | 22.2 | u.l. |
| 1.26 | 6480 | MITSUME | g | 20.5 | u.l. |
| 1.26 | 6480 | MITSUME | $R_c$ | 20.8 | u.l. |
| 1.26 | 6480 | MITSUME | $I_c$ | 20.4 | u.l. |
| 1.41 | 900 | DOT | R | 22.54 | 0.09 |
| 1.42 | 900 | DOT | I | 22.1 | 0.15 |
| 1.43 | 1200 | DOT | V | 23.09 | 0.19 |
| 2.7 | 2550 | CAHA | i' | 24.51 | 0.28 |
| 2.9 | 4775 | UVOT | m2 | 23.5 | u.l. |
| 3.4 | 3600 | DOT | R | 24.4 | u.l. |
| 4.0 | 900 | Gemini[100] | K | 22.4 | 0.1 |
| 4.4 | 3900 | DOT | R | 25.2 | u.l |
| 6.9 | 3780 | MMT[101] | K | 23.9 | 0.3 |
| 20.6 | 4000 | CAHA | i' | 23.2 | u.l. |
| 20.7 | 3200 | CAHA | r' | 24.4 | u.l. |
| 20.7 | 800 | CAHA | g' | 23.6 | u.l. |
| 22.5 | 7200 | DFOT | R | 24 | u.l. |
| Host Galaxy | | | | | |
| 142 | 1495 | UVOT | w2 | 21.64 | 0.14 |
| 2.9 | 4775 | UVOT | m2 | 21.52 | 0.15 |
| 140 | 2420 | UVOT | w1 | 21.50 | 0.18 |
| 0.2 | 877 | UVOT | u | 21.08 | 0.18 |
| 0.5 | 200 | DOT | B | 20.14 | 0.10 |
| 1.4 | 200 | DOT | V | 19.60 | 0.10 |
| 3.4 | 200 | DOT | R | 19.46 | 0.05 |
| 22.5 | 7200 | DFOT | R | 19.50 | 0.10 |
| 131 | 1400 | LDT | u | 20.95 | 0.04 |
| 172 | 300 | LDT | g | 19.93 | 0.01 |
| 172 | 300 | LDT | r | 19.50 | 0.01 |
| 131 | 600 | LDT | i | 19.20 | 0.01 |
| 131 | 600 | LDT | z | 19.07 | 0.02 |
| 111 | 2160 | HST | F814W | 19.24 | 0.01 |
| 114 | 4823 | HST | F160W | 18.93 | 0.01 |
| Archival | – | WISE[102] | W1 | 19.88 | 0.05 |
| Archival | – | WISE | W2 | 20.68 | 0.19 |

**Additional References**
77. Evans, P. A. et al., Methods and results of an automatic analysis of a complete sample of Swift-XRT observations of GRBs. *Mon. Not. R. Astron. Soc.* **397**, 1177-1201 (2009)
78. Breeveld, A. A. et al., Further calibration of the Swift ultraviolet/optical telescope. *Mon. Not. R. Astron. Soc.* **406**, 1687-1700 (2010)
79. Kotani, T. et al. MITSuME---Multicolor Imaging Telescopes for Survey and Monstrous Explosions. *Nuovo Cimento C Geophysics Space Physics C.* **28**, 755 (2005)
80. Yatsu, Y. et al. Development of MITSuME—Multicolor imaging telescopes for survey and monstrous explosions. *Physica E Low-Dimensional Systems and Nanostructures.* **40**, 434-437 (2007)
81. Shimokawabe, T. et al. MITSuME: multicolor optical/NIR telescopes for GRB afterglows. *AIP Conf. Ser.* **1000**, 543-546 (2008)
82. Niwano, M. et al. A GPU-accelerated image reduction pipeline. *Publications of the Astronomical Society of Japan.* **73**, 14-24 (2021)
83. Becker, A. HOTPANTS: High Order Transform of PSF ANd Template Subtraction. Astrophysics Source Code Library. ascl:1504.004 (2015)
84. Kumar, B. et al. 3.6-m Devasthal Optical Telescope Project: Completion and first results. *Bulletin de la Societe Royale des Sciences de Liege.* **87**, 29-41 (2018)
85. Pandey, S. B. et al. First-light instrument for the 3.6-m Devasthal Optical Telescope: 4Kx4K CCD Imager. *Bulletin de la Societe Royale des Sciences de Liege.* **87**, 42-57 (2018)
86. Kumar, A. et al. Photometric calibrations and characterization of the 4Kx4K CCD Imager, the first-light axial port instrument for the 3.6m DOT. Preprint at https://arxiv.org/abs/2111.13018 (2021)
87. Gupta, R. et al., GRB 211211A: Observations with the 3.6m Devasthal Optical Telescope. *GCN Circ.* 31299 (2021)
88. Bertin, E. & Arnouts, S. SExtractor: software for source extraction. *Astron. Astrophys. Suppl. Ser.* **117**, 393–404 (1996)
89. Ahumada, R. et al. The 16th Data Release of the Sloan Digital Sky Surveys: First Release from the APOGEE-2 Southern Survey and Full Release of eBOSS Spectra. *Astrophys. J. Sup. Ser.* **249**, 3 (2020)
90. Jordi, K., Grebel, E. K. & Ammon, K. Empirical color transformations between SDSS photometry and other photometric systems. *Astron. Astrophys.* **460**, 339-347 (2006)
91. Gonzaga, S., Hack, W., Fruchter, A. & Mack, J. *The DrizzlePac Handbook* (STScI, Baltimore, 2012)
92. Dong, Y.-Z., Gu, W.-M., Liu, T. & Wang, J. et al. A black hole-white dwarf compact binary model for long gamma-ray bursts without supernova association. *Mon. Not. R. Astron. Soc.* **475**, L101-L105 (2018)
93. Caito, L. et al. GRB060614: a ``fake'' short GRB from a merging binary system. *Astron. Astrophys.* **498**, 501-507 (2009)
94. Fryer, C. L., Woosley, S. E. & Hartmann, D. H. Formation Rates of Black Hole Accretion Disk Gamma-Ray Bursts. *Astrophys. J.* **526**, 152-177 (1999)
95. Gillanders, J. H., Sim, S. A. & Smartt, S. J. AT2018kzr: the merger of an oxygen-neon white dwarf and a neutron star or black hole. Mon. Not. R. Astron. Soc. **497**, 246-262 (2020)

96. Bobrick, A. et al. Transients from ONe white dwarf - neutron star/black hole mergers. *Mon. Not. R. Astron. Soc*. **510**, 3758-3777 (2022)
97. Gehrels, N. Confidence Limits for Small Numbers of Events in Astrophysical Data. *Astrophys. J.* **303**, 336 (1986)
98. Dichiara, S. et al. Short gamma-ray bursts within 200 Mpc. *Mon. Not. R. Astron. Soc.* **492**, 5011-5022 (2020)
99. Schlafly, E. F. & Finkbeiner, D. P. Measuring Reddening with Sloan Digital Sky Survey Stellar Spectra and Recalibrating SFD. *Astrophys. J.* **737**, 103 (2011)
100. Levan, A., et al. GRB 211211A - Gemini K-band detection. *GCN Circ.* **31235** (2021)
101. Rastinejad, J. et al. MMT/MMIRS Observations Indicate Fading of K-band Source. *GCN Circ.* **31264** (2021)
102. Marocco F. et al. The CatWISE2020 Catalog. *Astrophys. J. Sup. Ser.*, **253**, 8 (2021)